\documentclass[12pt,preprint]{aastex}

\newcommand{\p}{\partial}

\newcommand{\dd}{{\rm d}}
\newcommand{\ie}{i.e. }

\newcommand{\etal}{et al. }

\newcommand{\M}{{\cal M}}
\newcommand{\rsh}{{r}_{\rm sh}}
\newcommand{\sh}{{\rm sh}}
\newcommand{\Mc}{{{\cal M}^2}}
\newcommand{\z}{z}
\newcommand{\LL}{{\bar{\cal L}}}

\shorttitle{Convection versus advection}
\shortauthors{Foglizzo et al.}

\begin{document}

\title{Neutrino-driven convection versus advection in core collapse supernovae}
\author{T. Foglizzo}
\affil {Service d'Astrophysique, DSM/DAPNIA, UMR AIM CEA-CNRS-Univ. Paris VII,
CEA-Saclay, 91191 France}
\email{foglizzo@cea.fr}

\author{L. Scheck and H.-Th. Janka}
\affil{Max-Planck-Institut f\"ur Astrophysik, Karl-Schwarzschild-Str. 1, D-85741 Garching, Germany}

\author{ApJ received 2005 December 27; accepted 2006 August 16}

\begin{abstract}
A toy model is analyzed in order to evaluate the linear stability of the gain region immediately
behind a stalled accretion shock, after core bounce. This model demonstrates that a negative 
entropy gradient is not sufficient to warrant  linear instability. The stability criterion is governed by the ratio 
$\chi$ of the advection time through the gain region divided by the local timescale of buoyancy. The 
gain region is linearly stable if $\chi< 3$. The classical convective instability is recovered in the limit 
$\chi\gg3$. 
For $\chi>3$, perturbations are unstable in a limited range of horizontal wavelengths centered around 
twice the vertical size $H$ of the gain region. The threshold horizontal wavenumbers $k_{\rm min}$ 
and $k_{\rm max}$ follow simple scaling laws such that $Hk_{\rm min}\propto 1/{\chi}$ and 
$Hk_{\rm max}\propto\chi$. The convective stability of the $l=1$ mode in 
spherical accretion is discussed, in relation with the asymmetric explosion of core collapse supernovae. 
The advective stabilization of long wavelength perturbations weakens the possible influence of convection alone on a global $l=1$ mode. 
\end{abstract}


\section{Introduction}

Convective instabilities may be an important ingredient of the explosion mechanism of core collapse 
supernovae. Below the neutrinosphere, they can increase the neutrino luminosity, and in the neutrino 
heating layer they can help pushing the shock farther out. Convection in the supernova core may also
be the seed for the large-scale anisotropies seen in many supernovae and supernova remnants and
might be linked to the measured high velocities of young pulsars (e.g., Arnett 1987, Woosley 1987,
Herant et al.\ 1992).
Negative gradients of entropy were 
initially thought to arise as a natural consequence of the decline of the shock strength due to 
photodissociation of heavy nuclei and neutrino escape 
(Arnett 1987, Burrows 1987, Bethe, Brown \& Cooperstein 
1987, Bethe 1990). A more durable effect was recognized by Herant, Benz \& Colgate (1992) in their 
simulations: neutrino heating is able to maintain a negative entropy gradient in a ``gain region" 
immediately behind the stalled shock. They also observed that the convective eddies tend to merge 
and produce eddies of the size of the computing box. Similar results were found in the numerical 
simulations of Herant et al. (1994), Burrows et al. (1995), Janka \& M\"uller (1996), 
Mezzacappa et al. (1998).

Are such convective instabilities able to produce an $l=1$ asymmetry as
suggested by Herant (1995) and Thompson (2000) and seen more recently 
in numerical simulations (Scheck et al.\ 2004)?
Estimates of the linear growth rate and wavelength of the nonspherical modes found in these studies 
cannot be directly made on grounds of the considerations of convective instabilities in hydrostatic 
spherical shells (e.g., Chandrasekar 1961).
Attention has to be paid to the fact that the advection of matter across the shock and through the gain 
region might seriously reduce the convective growth rate and modify the spatial structure of unstable 
modes. This paper is dedicated to evaluating and characterizing the magnitude of this potentially stabilizing effect.
This question has become particularly acute since the discovery of another hydrodynamical mechanism which might be responsible for an $l=1$ asymmetry.
The nonspherical modes of deformation of an accretion shock discovered in adiabatic numerical 
simulations by Blondin \etal (2003), which the authors termed SASI---standing accretion shock
instability--- are independent of convection. This instability seems to be due to an advective-acoustic cycle, based on the acoustic feedback produced by the advection of entropy and 
vorticity perturbations from the shock to the accretor (Foglizzo \& Tagger 2000, Foglizzo 2001, 2002). 
More realistic simulations by Scheck \etal (2004), including neutrino heating, a microphysical
equation of state and the environment of collapsing stellar cores, recognized the development
of a strong $l=1$ mode possibly due to the combination of convective and advective-acoustic 
instabilities. The asymmetry produced by this instability makes it a good candidate to explain the high 
velocities of pulsars. The mechanism responsible for this instability is still a matter of debate, since Blondin \& Mezzacappa (2006) advocated a purely acoustic origin whereas Ohnishi \etal (2006) recognize an advective-acoustic cycle.

Is it possible to disentangle the convective from other instabilities from the point of 
view of their linear growth rates and spatial structure?
As a first step, the present study aims at a better characterization of neutrino-driven convection in the
gain layer beyond the classical hydrostatic approach.  
In order to distinguish convection in the gain region from any type of instability based on an acoustic feedback produced {\em below} the gain radius, we choose to analyze the onset of convection in a particular set up which neglects such an acoustic feedback. For this purpose we build in Sect.~3 a simple toy model incorporating the minimum ingredients leading to the convective instabilitiy below a stationary shock: a parallel flow in Cartesian geometry, in a uniform gravity. This flow is simple enough to allow for a full characterization of its stability properties (Sect.~4). The extrapolation of these properties, when convergence effects are included, is then tested by solving the same boundary value problem in spherical geometry (Sect.~5). This allows us to address the question of the convective destabilization of the $l=1$ mode during the phase of stalled shock of core collapse supernovae. 
The results of our perturbative approach are confronted to two examples of numerical simulations in Sect.~6, which illustrate two situations where the instabilities can be disentangled. 
Conclusions are drawn in Sect.~7. 
Before that, let us first recall the classical results concerning the convective instability in plane and spherical geometry.

\section{Classical results about the onset of convection in a hydrostatic equilibrium\label{Sectclassical}}

\begin{figure}
\plotone{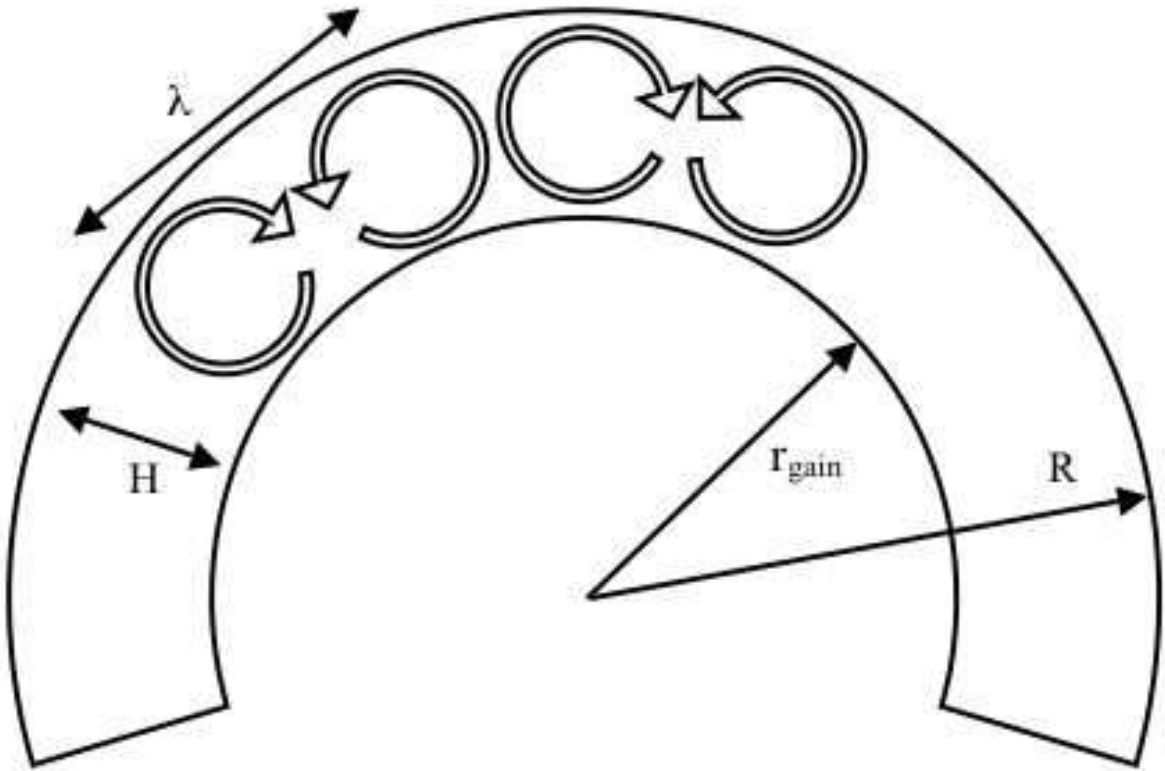}
\caption[]{Schematic view of convection in a spherical shell of size $H$. The first unstable modes when 
viscosity is decreased have a wavelength $\lambda\sim (2$--$3)H$, depending on the boundary conditions (Chandrasekhar 1961)}
\label{figconvsp}
\end{figure}

In the absence of viscosity and of stabilizing composition gradients, 
a stratified atmosphere with a negative entropy gradient is unstable at all 
wavelengths. Perturbations with a horizontal wavelength shorter than the scale height $H$ of the entropy gradient are the most unstable. 
In a perfect gas with an adiabatic index $\gamma$, a measure of the entropy is defined by
the dimensionless quantity $S$, as a function of pressure $P$ and density $\rho$:
\begin{eqnarray}
S&\equiv&{1\over\gamma-1}\log\left\lbrack{P\over P_{\rm sh}}
\left({\rho_{\rm sh}\over\rho}\right)^\gamma\right\rbrack.\label{defS}
\end{eqnarray}
In this formula, pressure and density are normalized by their value immediately after the shock. In what follows, the subscript ``sh" always refers to postshock quantities. The maximum growth rate 
${\omega}_{\rm buoy}$ is given by the Brunt-V\"ais\"ala frequency, expressed by the 
gravitational acceleration $G$ and $H$:
\begin{eqnarray}
{\omega}_{\rm buoy}&\equiv&G^{1\over 2}\left|{\nabla P\over \gamma P}
-{\nabla\rho\over\rho} \right|^{1\over 2}=\left({\gamma-1\over\gamma}G\nabla S\right)^{1\over2},
\nonumber\\
&\sim& \left({G\over H}\right)^{1\over 2}\label{wconv}.
\end{eqnarray}
Perturbations with a longer horizontal wavelength than $H$ are also unstable, with a slower growth 
rate however.
Perturbations with a horizontal wavelength much shorter than $H$ are easily stabilized by a small 
amount of viscosity. This is illustrated by the calculations of Chandrasekhar (1961) of the onset of 
convection, either between two parallel plates or in a spherical shell. These calculations measured the 
amount of viscosity which is required to stabilize a perturbation with a given wavelength. The 
wavelength of the first unstable mode is about $[2-3]$ times the vertical size of the unstable region 
depending on the nature of the boundaries (free, rigid or mixed). Note that a factor 2 would be rather 
intuitive, since it corresponds to a pair of two counter-rotating circular eddies (see 
Fig.~\ref{figconvsp}).
In a spherical shell, a naive estimate of the azimuthal number $l$ of the first unstable perturbations, 
based on the number of pairs of circular eddies which would fit in the unstable shell $r_{\rm 
gain}<r<R$, leads to:
\begin{eqnarray}
l\sim{\pi\over2}{R+r_{\rm gain}\over H}\ .
\end{eqnarray}
This simplistic approach is compatible with the exact calculations performed by Chandrasekhar (1961), 
within the same factor $1-2$ as in the case of Benard convection (Rayleigh 1916). This factor depends 
on the boundary conditions, on the gravity profile, and can be interpreted as an aspect ratio of the 
eddies, which are not circular. A direct application to the size of a stalled shock with $R\sim 150$ km 
and $r_{\rm gain}\sim 100$ km, as in Herant, Benz \& Colgate (1992), would lead to $l\sim 7$. 
As noted by Herant, Benz \& Colgate (1992), the increase of $H$ naturally leads to the decrease 
of the optimal $l$.
Is the residual instability of the $l=1$ mode fast enough to have a significant influence during the first 
second after core bounce? The classical description by Chandrasekhar is not directly applicable 
here, not only because viscosity is negligible, but also because it does not take into account the 
presence of a shock wave, and the associated flow of gas across it. Let us compare the timescale of 
buoyancy $\omega_{\rm buoy}^{-1}$ with the advection timescale $H/v_{\rm sh}$ through the gain region. The local gravity at the shock radius $r_{\rm sh}$ is 
$G\equiv {\cal G}M/r_{\rm sh}^2$, where ${\cal G}$ is the gravitational constant and $M$ is the enclosed gravitating mass. In what follows, the subscript ``1" refers to preshock quantities. Assuming that the gas is in free fall ahead of the shock ($v_1^2\sim 2{\cal G}M/r_{\rm sh}$), one estimates:
\begin{eqnarray}
{H\omega_{\rm buoy}\over v_{\rm sh}}&\sim&
\left({{\cal G}M\over r_{\rm sh}v_{\rm sh}^2}\right)^{1\over 2}
\left({H\over r_{\rm sh}}\right)^{1\over 2}
,\\
&\sim&3.1\left({v_1\over 7v_{\rm sh}}\right)\left({H\over 0.4r_{\rm sh}}\right)^{1\over 2}.\label{Estichi}
\end{eqnarray}
A typical radius of the stalled shock 
is $r_{\rm sh}\sim 150$ km, and $0.3\le H/r_{\rm sh}\le 0.5$. 
The velocity jump across the shock would be $v_1/v_{\rm sh}=7$ for an adiabatic gas with $\gamma=4/3$. 
This ratio may increase up to $v_1/v_{\rm sh}\sim 10$ due to dissociation of iron into nucleons. 
Even then, the rough estimate of Eq.~(\ref{Estichi}) indicates that the convective growth time is 
comparable to the advection time through the gain region.

\section{Description of a planar toy model}

\begin{figure}
\plotone{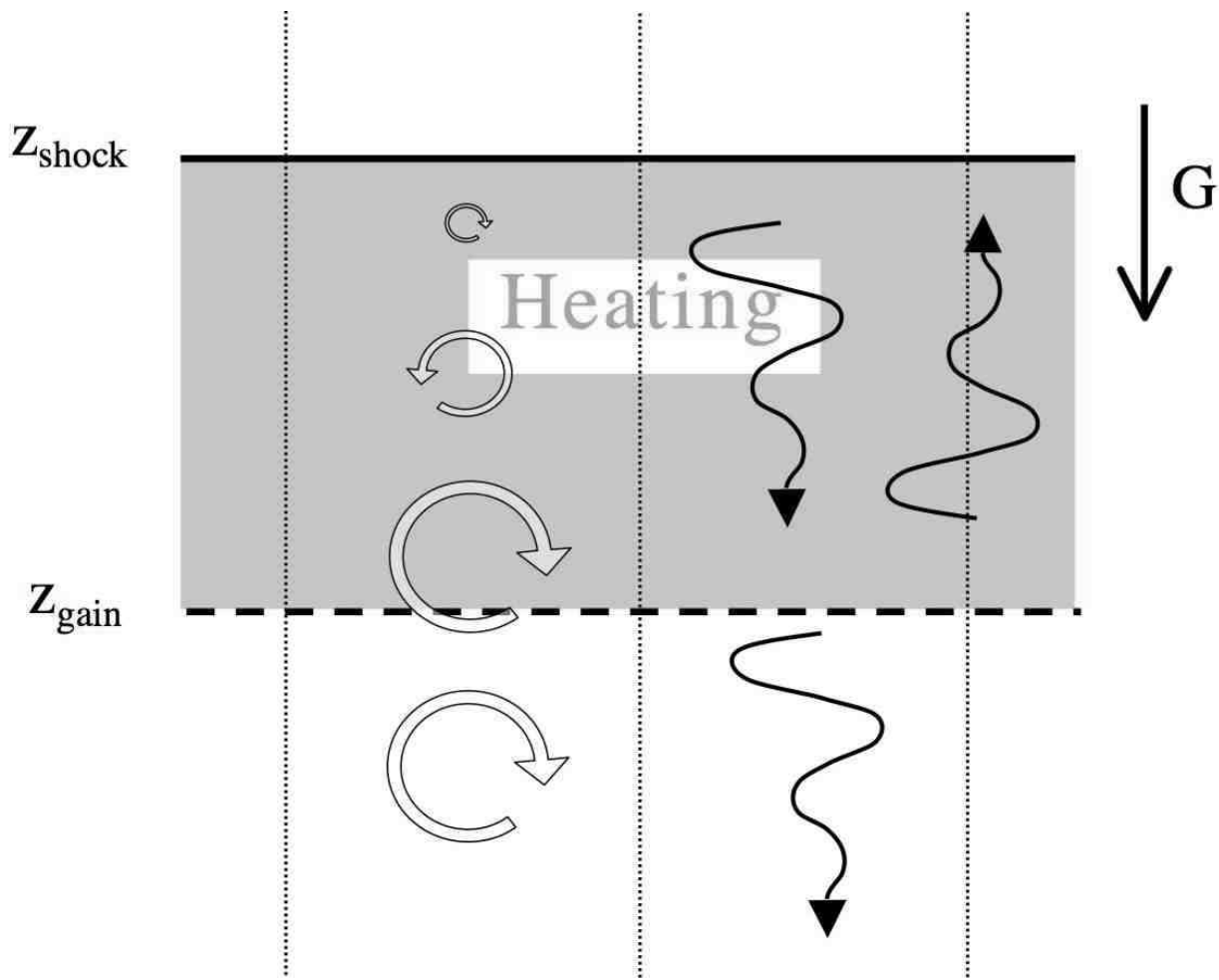}
\caption[]{Schematic view of the toy model enabling the convective instability  
immediately below a shock. Heating dominates cooling in the gain region, 
so that the entropy gradient is oriented downward. Gravity $G$ is uniform.  
The flow below the gain radius is assumed to be uniform in the 
stationary regime, in order to preclude any acoustic feedback from below 
once perturbed. Entropy/vorticity perturbations (circular arrows) are 
advected downward with the flow. Acoustic perturbations  (wavy arrows) 
propagate downward below the gain radius.}
\label{figtoy}
\end{figure}

This Section establishes the equations describing a toy model in Cartesian 
geometry, illustrated by Fig.~\ref{figtoy}, in which only the minimum ingredients 
leading to the convective instability have been included. 
This toy model mimics in a most simplified form the accretion flow in the gain 
region immediately below the stalled accretion shock, a few tens of 
milliseconds after core bounce. 
The eigenmodes are solved numerically in Sect.~\ref{convadv}.

\subsection{Stationary flow}

\subsubsection{General description}

The stationary flow is parallel along the $z$ direction, 
in a uniform gravity $G$. Self-gravity is neglected. A shock is stationary at the height $z_{\rm sh}$. 
The flow is described by a perfect gas with an adiabatic index $\gamma=4/3$, 
corresponding to a gas of relativistic electrons or photons and 
electron-positron pairs. We shall further assume that $P\propto \rho T$ (with $T$
being the temperature), which is a suitable description 
of the thermodynamic conditions in the gain layer independent
of whether relativistic particles or baryons dominate the
pressure (for details, see Janka 2001, Bethe 1993).
The essential ingredients of the convective instability are the entropy changes 
and the local acceleration. In the gain region, the local acceleration is mainly 
due to the gravity $G$. The entropy gradient ${\nabla S}$ is produced 
by the heating through neutrino absorption, which exceeds the cooling by 
neutrino emission in the gain region. The heating/cooling function ${\cal L}$ 
is adapted from Bethe \& Wilson (1985), neglecting the effect of geometric 
dilution of the neutrino flux:
\begin{eqnarray}
{\cal L}&\equiv&{\LL\rho\over 1-\beta}\left\lbrack1-
\beta\left({T\over T_{\rm sh}}\right)^6\right\rbrack,\label{heatfunc}\\
\LL&\sim&2.2\left({r_{\rm sh}\over 150{\rm km}}\right)^{-2}
\times10^{20}{\rm ergs\;g^{-1}\;s^{-1}},\label{realL}
\end{eqnarray} 
where $T_{\rm sh}$ is the postshock temperature. The parameter $\beta<1$ is defined 
as the ratio of the strengths of 
neutrino cooling and neutrino heating at the shock. The gain radius 
$z_{\rm gain}$ is defined as the point where ${\cal L}=0$. According to 
Eq.~(\ref{heatfunc}), the temperature at the gain radius is 
$T_{\rm gain}=T_{\rm sh}/\beta^{1\over6}$. The temperature contrast 
$\Delta T_{\rm gain}/T_{\rm sh}$ within the gain region
is thus directly related to the parameter  $\beta$ of the heating/cooling function through
\begin{eqnarray}
{\Delta T_{\rm gain}\over T_{\rm sh}}\equiv {T_{\rm gain}-T_{\rm sh}\over T_{\rm sh}}
={1\over\beta^{1\over6}}-1.\label{betat}
\end{eqnarray}
Heating and cooling are neglected above and below the gain region. \\
The equation of continuity, the Euler equation and the entropy equation defining the stationary flow 
in the gain region lead to the following differential system:
\begin{eqnarray}
{\p \over\p \z}(\rho v) &=&0\ ,\label{eqcont}\\
{\p\over\p \z}\left({v^2\over 2}+{c^2\over\gamma-1}+G\z\right)&=&
{{\cal L}\over \rho v}\ ,\label{eqbern}\\
{\p S\over \p \z}&=&{{\cal L}\over Pv}\ ,\label{eqS}
\end{eqnarray}
where the last relation makes use of $P\propto \rho T$, and the pressure force in the Euler equation has been transformed using the definition (\ref{defS}) of $S$ and the relation $c^2\equiv \gamma P/\rho$:\begin{eqnarray}
{\nabla P\over\rho}=\nabla\left( {c^2\over\gamma-1}\right)-{c^2\over\gamma}\nabla S.\label{gradP}
\end{eqnarray}

\subsubsection{Photodissociation at the shock}

Although dissociation at the shock is not expected to be an important ingredient for the convective instability, it is taken into account because of its effect on the postshock velocity, and thus on the advection time through the gain region. The effect of dissociation can be crudely incorporated by assuming that it takes place immediately behind the shock. It is parametrized in Appendix~A by a decrease of the postshock Mach number $\M_{\rm sh}$ below the adiabatic value $\M_{\rm ad}$: 
\begin{eqnarray}
\M_{\rm sh}\le\M_{\rm ad}\equiv\left\lbrack{2+(\gamma-1)\M_1^2\over2\gamma\M_1^2-
\gamma+1}\right\rbrack^{1\over2}.
\end{eqnarray}
Note that Mach number are defined as positive $\M\equiv |v|/c>0$.
In the phase of a stalled accretion shock the gravitating mass is $M\sim1.2\;M_{\sun}$ and 
the preshock sound speed is $c_1\sim8.5\times10^8{\rm cm\;s^{-1}}$.
The incident Mach number $\M_1$ is estimated assuming free-fall velocity of the gas incident at the shock:
\begin{eqnarray}
|v_1|&\sim&\left({2{\cal G}M\over r_{\rm sh}}\right)^{1\over2},\nonumber\\
&\sim&4.6\left({r_{\rm sh}\over 150{\rm km}}\right)^{-{1\over2}}\times 10^9{\rm cm\;s^{-1}},\label{realv1}\\
\M_1&\sim&5.4\left({r_{\rm sh}\over 150{\rm km}}\right)^{-{1\over2}}.
\end{eqnarray}
The incident Mach number is taken equal to $\M_1=5$ in the rest of the paper, so that 
$\M_{\rm ad}\sim0.39$ if $\gamma=4/3$. A prescription $\M_{\rm sh}/\M_{\rm ad}=0.77$ accounting for dissociation 
corresponds to more realistic values of $\M_{\rm sh}\sim0.3$, a velocity jump $v_1/v_{\rm sh}\sim 8.8$ 
and a jump of sound speed $c_{\rm sh}/c_1\sim1.85$. 
Considering the full range $1/(\gamma\M_1)<\M_{\rm sh}\le\M_{\rm ad}$ allows us to explore the stability of a larger family of flows, and 
thus better understand the onset of convection. The extreme case $\M_{\rm sh}\M_1=1/\gamma$ corresponds 
to $c_{\rm sh}=c_1$ and $v_1/v_{\rm sh}=\gamma\M_1^2$ is refered to as the ``isenthalpic shock".

\subsubsection{Dimensionless parameters of the toy model}

\begin{figure}
\plotone{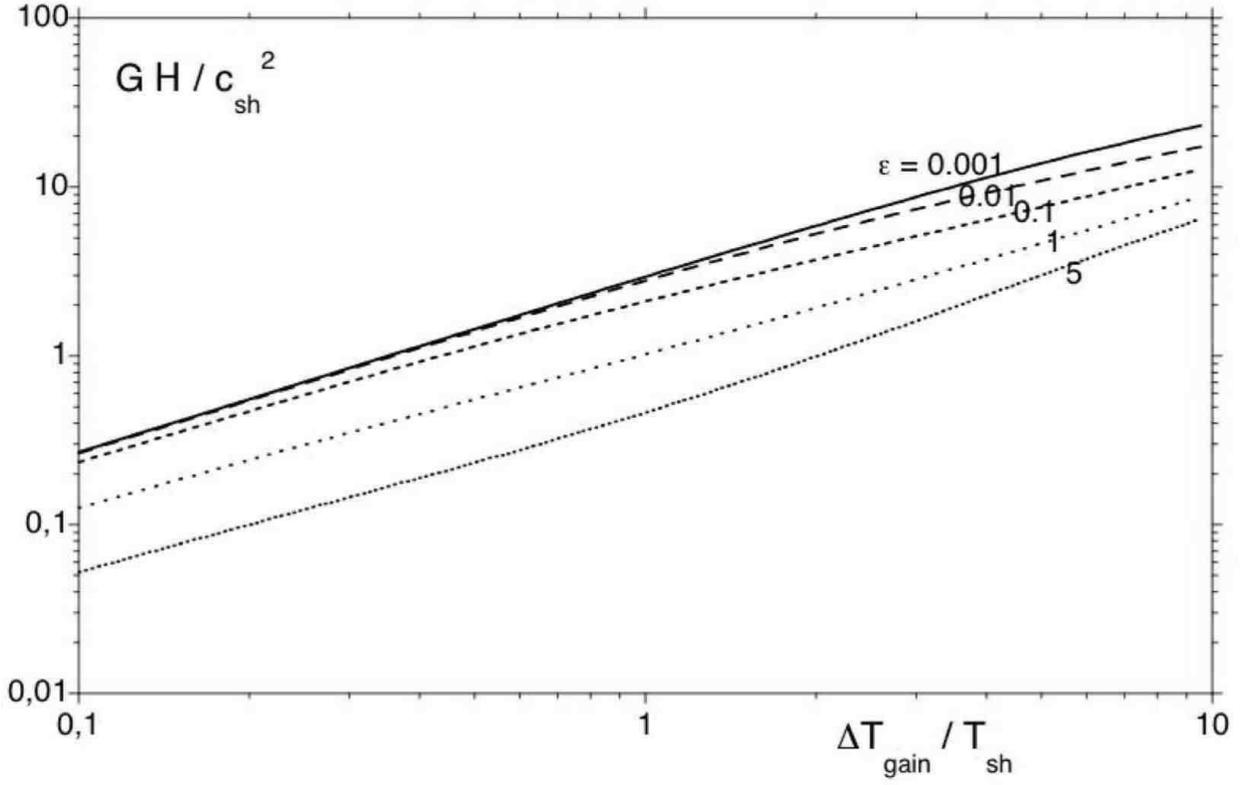}
\caption[]{The power of heating, cooling and gravity are expressed through the dimensionless parameters $\epsilon$ and $\Delta T_{\rm gain}/T_{\rm sh}$. The resulting size 
$H\equiv \z_{\rm sh}-\z_{\rm gain}$ of the gain region is measured in units of $c_{\rm sh}^2/G$. The shock is adiabatic ($\M_1=5$ in all plots).
}
\label{figgdr}
\end{figure}
The heating constant $\LL>0$ can be compared to the critical heating rate $\LL_{\rm crit}$ needed to cancel the velocity gradient immediately behind the shock, as deduced from Eqs.~(\ref{eqcont}-\ref{eqS}): 
\begin{eqnarray}
{1\over v}{\p v\over\p z}&=&{1\over c^2-v^2}\left\lbrack G+(\gamma-1)
{{\cal L}\over\rho v}\right\rbrack,\label{dvdz}\\
\LL_{\rm crit}&=&-{G v_\sh\over\gamma-1}.\label{defH0}
\end{eqnarray}
The dimensionless parameter defined by $\epsilon\equiv \LL/\LL_{\rm crit}$ is thus a measure of the power of heating from the dynamical point of view. The flow is decelerated after the shock if $\epsilon<1$. Using a gravitating mass 
$M=1.2\;M_{\sun}$ in Eq.~(\ref{defH0}) and measuring the post-shock velocity in units of the free fall velocity $v_1$ (Eqs.~\ref{realL} and \ref{realv1}), a typical value of $\epsilon$ is:
\begin{eqnarray}
\epsilon&\sim& 0.16\;\left({v_1\over7v_{\rm sh}}\right)
\left({r_{\rm sh}\over 150{\rm km}}\right)^{1\over 2} .\label{estiepsilon}
\end{eqnarray}
In addition to $\gamma\sim 4/3$ and $\M_1\sim 5$, the independent parameters of this toy 
model are the dissociation parameter $\M_{\rm sh}/\M_{\rm ad}$, a measure of the heating 
rate at the shock through the parameter $\epsilon$, and the temperature contrast 
$\Delta T_{\rm gain}/T_{\rm sh}$ within the gain region.  The parameters of this toy model allow for a wide variation of the size $H\equiv \z_{\rm sh}-\z_{\rm gain}$ of the gain region. This can be seen by integrating the differential system Eqs.~(\ref{eqcont}-\ref{eqS}) from the shock to the gain radius, 
with the function ${\cal L}$ described by Eq.~(\ref{heatfunc}).
The corresponding results for a flow without dissociation are displayed in Fig.~\ref{figgdr}.

\subsection{Linear perturbations}

\subsubsection{Differential system}

The 1-D stationary flow is perturbed in the plane $x,z$, where $x$ is the horizontal direction. 
The complex frequency $\omega\equiv(\omega_r,\omega_i)$ of the perturbation is defined such that 
the real part $\omega_r$ is the oscillation frequency, and the imaginary part $\omega_i$ is the growth rate. The eigenmodes calculated in this study are non oscillatory ($\omega_r=0$).
Rather than $\delta v_x$, $\delta v_z$, $\delta c$ 
and $\delta\rho$, the functions chosen for the perturbative approach are the entropy $\delta S$, and three functions $f$, $ h$ and $\delta K$, defined by:
\begin{eqnarray}
f&\equiv&v\delta v_z+{2\over\gamma-1}c\delta c\, ,\label{defif}\\
 h&\equiv&{\delta v_z\over v}+{\delta\rho\over\rho},\label{defih}\\
\delta  K&\equiv& {iv k_x}\delta w_y+{k_x^2c^2\over\gamma}
\delta S,\label{defK1}
\end{eqnarray}
where $k_x$ is the horizontal wavenumber and $\delta w_y$ is the perturbation of vorticity.
This choice is motivated by the fact that in the adiabatic limit, the perturbations $\delta S$ and 
$\delta K$ are conserved when advected (Foglizzo 2001, hereafter F01) and the coefficients of the differential 
system expressed with $f,h$ contain no radial derivative of the stationary flow quantities $v,c,\M$. 
If $\omega\ne 0$, the linearized equations are expressed by the following differential system of fourth order: 
\begin{eqnarray}
{\p f\over\p \z}={i\omega v\over 1-\Mc}\left\lbrace
 h -{f\over c^2} 
 +
\left\lbrack\gamma-1+{1\over\Mc}\right\rbrack{\delta S\over\gamma}
 \right\rbrace\nonumber\\
 +\delta\left({{\cal L}\over \rho v}\right), \label{dfp}
\\
{\p{  h}\over\p \z}={i\omega\over v(1-\Mc)}\left\lbrace
\frac{\mu^{2} }{c^{2}} f
 -\Mc {  h}
- \delta S\right\rbrace\nonumber\\
+{i\delta K\over\omega v}, \label{dhp}
\\
{\p \delta S\over\p \z}={i\omega\over v}\delta S
+\delta\left({{\cal L}\over pv}\right), \label{dsp}
\\
{\p\delta K\over \p \z}={i\omega\over v}\delta K
+k_x^2\delta\left({{\cal L}\over \rho v}\right), \label{dkp}
\end{eqnarray}
where $\mu$ is defined by:
\begin{eqnarray}
\mu^2\equiv1-{k_x^2c^2\over\omega^2}(1-\Mc).
\end{eqnarray}
$\mu$ is a natural parameter in the algebraic formulation of the problem. 
When the frequency $\omega$ is real, $\mu$ is directly related to the angle $\psi$ between 
the direction of propagation of the wave and the direction of the flow. Correcting a typing error in 
Eq.~(E11) of F02:
\begin{eqnarray}
\tan^2\psi &=&{1-\mu^2\over\mu^2(1-\M^2)}.
\end{eqnarray}
In the differential system (\ref{dfp}-\ref{dkp}), the perturbations $\delta({\cal L}/\rho v)$ and $\delta({\cal L}/P v)$ can be expressed in terms of $f,h,\delta S$ using Eqs.~(\ref{defif}-\ref{defih}) and a perturbation of Eq.~(\ref{heatfunc}) (see Appendix~B).

\subsubsection{Boundary condition at the shock}

The boundary conditions at the shock surface are obtained in Appendix~C.1 using conservation laws in the frame of the 
perturbed shock:
\begin{eqnarray}
f_{\rm sh}&=&\Delta v (v_{\rm sh}-v_{1})-\Delta \zeta
\frac{c_\sh^{2}}{\gamma} \nabla{S}_{\rm sh} ,\label{fsh}\\
 h_{\rm sh}&=&{\Delta v\over v_{\rm sh}} \left(1-{v_{\rm sh}\over v_{1}}\right) ,\label{hsh}\\
{\delta S_{\rm sh}\over\gamma}&=&-\Delta \zeta
\left\lbrack {\nabla S_{\rm sh}\over\gamma}+
\left(1-{v_{\rm sh}\over v_{1}}\right){G\over c_\sh^{2}}
\right\rbrack\nonumber\\
&&-{ v_1\Delta v\over c_\sh^{2}}\left(1-{v_{\rm sh}\over v_1}\right)^{2}, \label{Ssh}\\
\delta K_{\rm sh}&=&
-k_x^2\Delta \zeta
\frac{c_\sh^{2}}{\gamma} \nabla{S}_{\rm sh}\ ,\label{Ksh}
\end{eqnarray}
where the velocity of the shock is related to its displacement through 
$\Delta v\equiv -i\omega\Delta\zeta$. In these equations, the cooling/heating above the shock is 
neglected (${\cal L}_1\ll{\cal L}_{\rm sh}$). 
These boundary conditions agree with FGR05 when heating is suppressed.

\subsubsection{Leaking condition at the lower boundary}

The effect of negative entropy gradients within the gain region are separated
from any coupling process below the gain radius by choosing a leaking 
boundary condition at the gain radius:
\par - entropy and vorticity perturbations reaching the gain radius are simply 
advected downward,
\par - acoustic perturbations are free to propagate downward, with no reflexion.\\
This boundary condition is equivalent to replacing the accretion flow below the gain
radius by a uniform flow. The uniformity of the unperturbed flow warrants the absence of coupling
processes, once perturbed. Each perturbation is decomposed in Appendix~C.2 as the sum of entropy, 
vorticity, and pressure perturbations:
\begin{eqnarray}
f&=&f_S+f_K+f_++f_-\ ,\label{fpm}\\
 h&=& h_S+ h_K+ h_++ h_-\ .\label{hpm}
\end{eqnarray}
The absence of coupling below the gain radius corresponds to the absence of an 
acoustic flux from below ($f_-=0$, $h_-=0$). According to the calculation of Appendix~C.2, 
this requirement is equivalent to the following condition:
\begin{eqnarray}
{\mu\over \M}{f\over c^2} -  h-\left(\gamma+{\mu\over\M}{1-\M^2\over1+\mu\M}\right)
{\delta S\over\gamma}\nonumber\\
+{1-\mu^2\over1+\mu\M}{\delta K\over k_x^2c^2}=0.\label{bcleaking}
\end{eqnarray}

\section{Convective mode in the gain region \label{convadv}}

\subsection{Definition of the ratio ${\chi}$ comparing the advective and buoyancy timescales}

\begin{figure}
\plotone{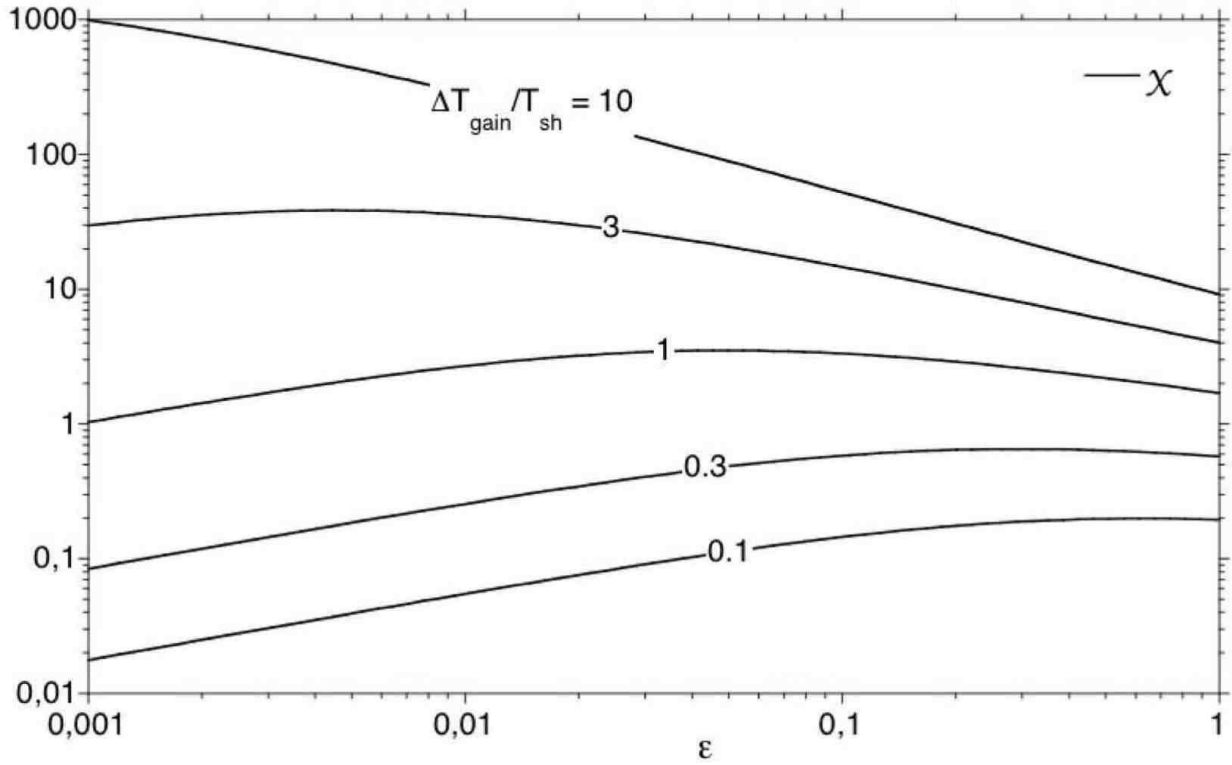}
\caption[]{Ratio ${\chi}$ of the advective and convective timescales in our toy model with an adiabatic 
shock, as function of the heating parameter $\epsilon$, for different values of 
$\Delta T_{\rm gain}/T_{\rm sh}$ indicated on each curve. }
\label{figtauRT}
\end{figure}

The maximum growth rate $\omega_{\rm buoy}$ of the convective instability (Eq.~\ref{wconv}) can be 
expressed with the local variables $\epsilon,T,v,c$ as a function of height $z$:
\begin{eqnarray}
{\omega}_{\rm buoy}(z)
&=&\epsilon^{1\over2}
\left|{v_{\rm sh}\over v}\right|^{1\over2}
\left\lbrack{1-(T/T_{\rm gain})^6\over 1-(T_{\rm sh}/T_{\rm gain})^6}
\right\rbrack^{1\over2}
{G\over c},\\
{\omega}_{\rm max}&\equiv&{\rm Max}_{z_{\rm gain}<z<z_{\rm sh}}{\omega}_{\rm buoy}(z).\label{wbuoymax}
\end{eqnarray}
Note that $\epsilon^{1\over2}$ directly measures the local growth rate of the convective instability at the 
shock, in units of $G/c_{\rm sh}$:
\begin{eqnarray}
{\omega}_{\rm buoy}^{\rm sh}
&=&\epsilon^{1\over2}
{G\over c_{\rm sh}}.
\end{eqnarray}
When considered as a local instability, the transient amplification of short wavelength perturbations, 
during their advection through the gain region, can be estimated by the quantity $\exp{\chi}$, with
\begin{eqnarray}
{\chi}&\equiv& \int_{{\rm gain}}^{{\rm shock}} {\omega}_{\rm buoy}(z) {\dd \z \over v}.\label{defT}
\end{eqnarray}
$\chi$ can be interpreted as the ratio of the advective timescale to some averaged timescale of convective growth.
The correspondence between $\epsilon$, $\Delta T_{\rm gain}/T_{\rm sh}$ and $\chi$ is shown in 
Fig.~\ref{figtauRT}. 

\subsection{Numerical solution of the eigenmode problem}

\begin{figure}
\plotone{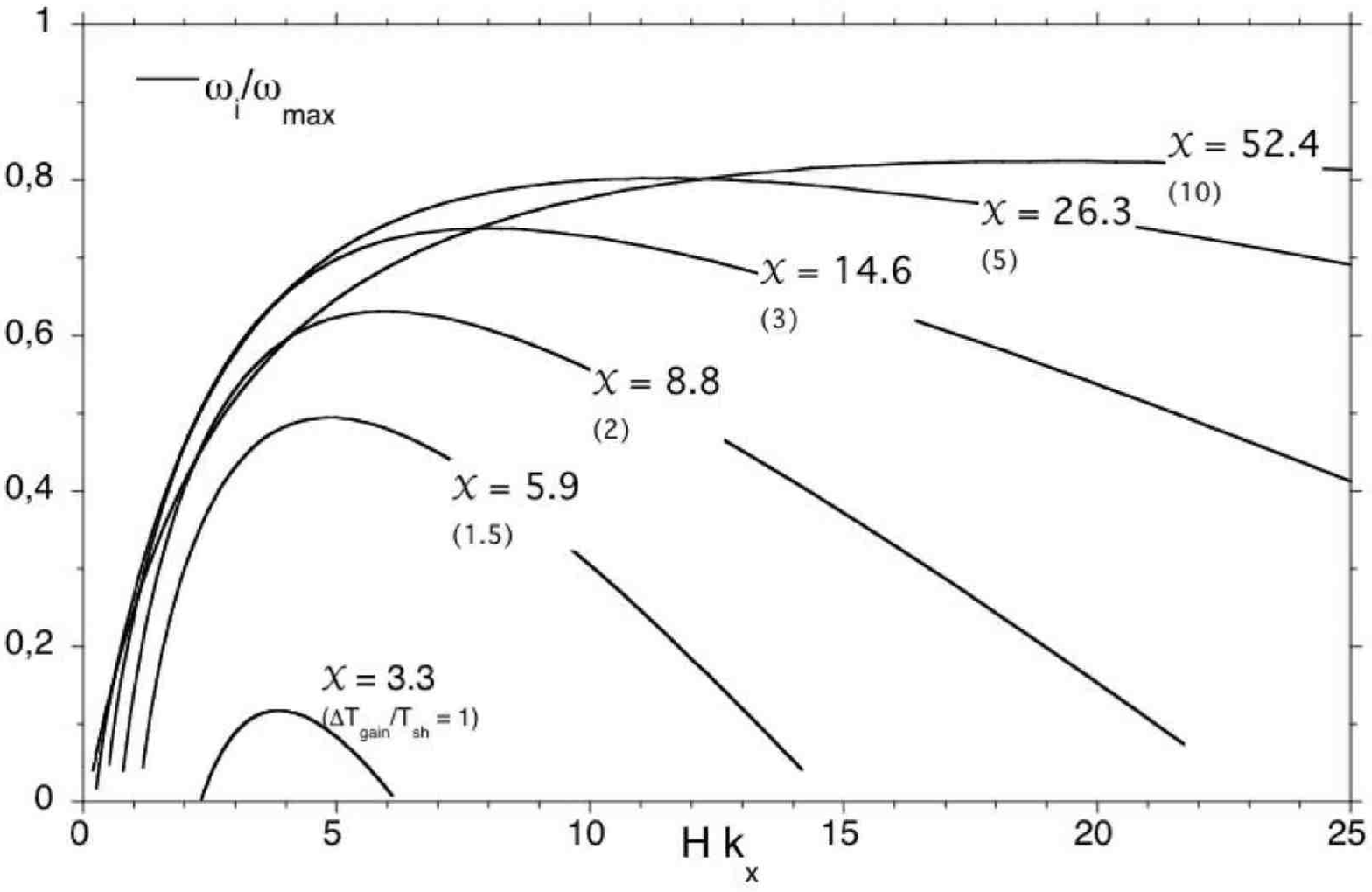}
\caption[]{Growth rate $\omega_i$ of the convective instability as a function of the horizontal wavenumber. Growth rates 
are normalized to $\omega_{\rm max}$. Heating is such that $\epsilon=0.1$.
The shock is adiabatic. The ratio ${\chi}$ of the advective and convective timescales is indicated on 
each curve, with the corresponding temperature contrast $\Delta T_{\rm gain}/T_{\rm sh}$ 
given in parentheses. Advection stabilizes both the long and short wavelengths. The 
convective instability disappears for ${\chi}<3$.}
\label{figconv}
\end{figure}
\begin{figure}
\plotone{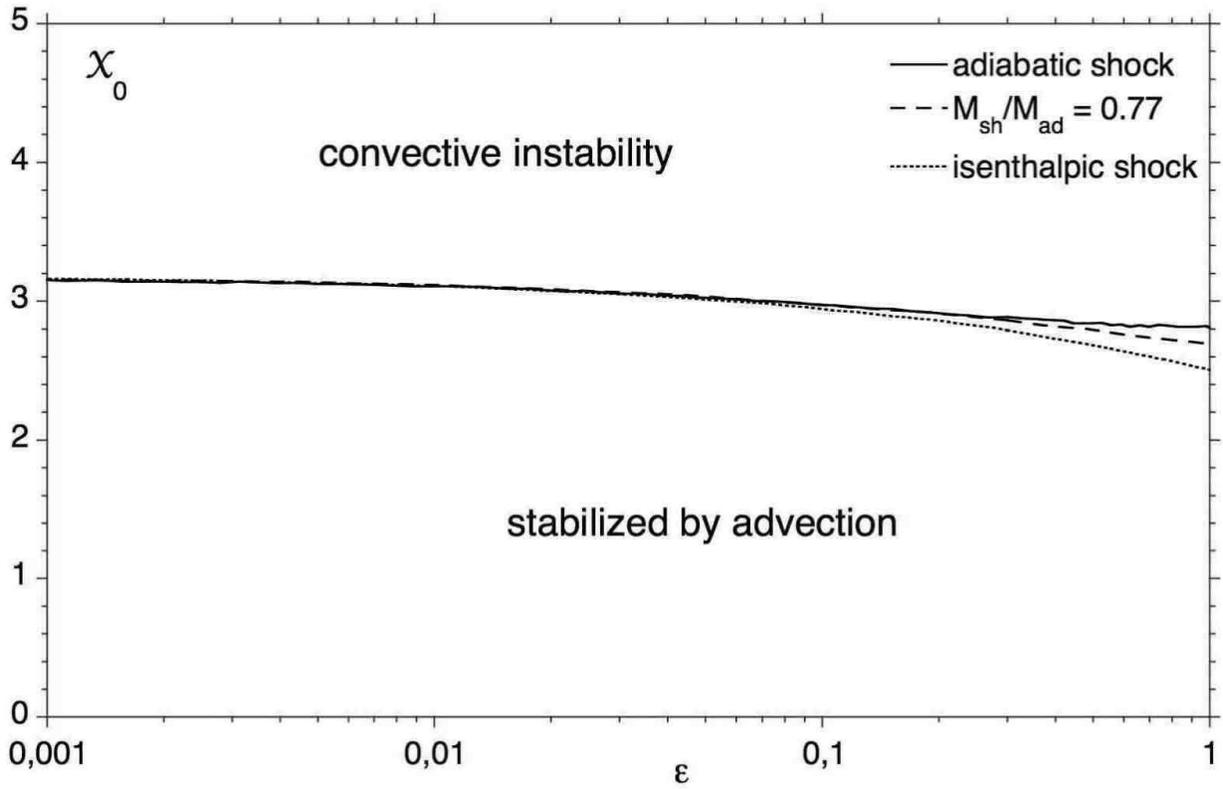}
\caption[]{Threshold ${\chi}_0$ of the ratio of the convective and advective timescales, determining the 
existence of the convective mode, as a function of the power $\epsilon$ of heating. This threshold is 
essentially insensitive to dissociation at the shock.}
\label{figthreshTau}
\end{figure}
\begin{figure}
\plotone{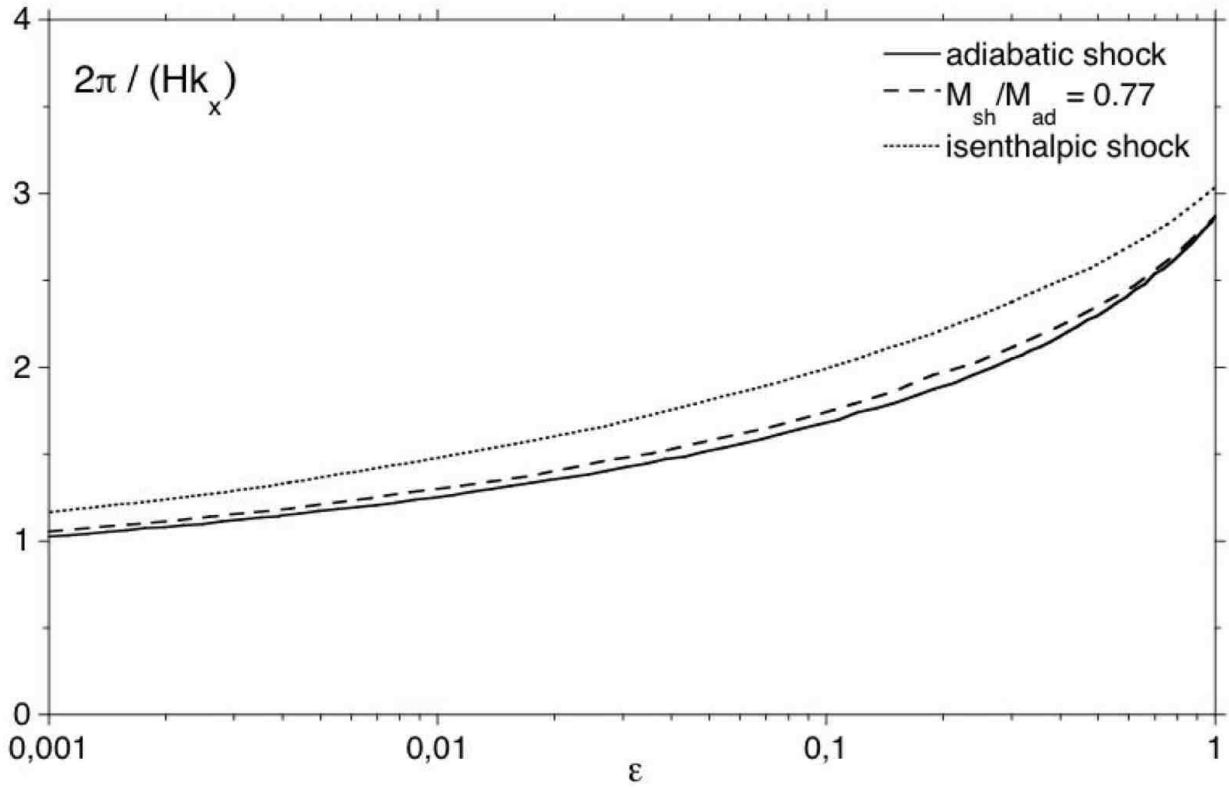}
\caption[]{
Wavelength of the neutral mode corresponding to the threshold value
$\chi_0$, in units of the size of the gain region.}
\label{figwavenumber}
\end{figure}
\begin{figure}
\plotone{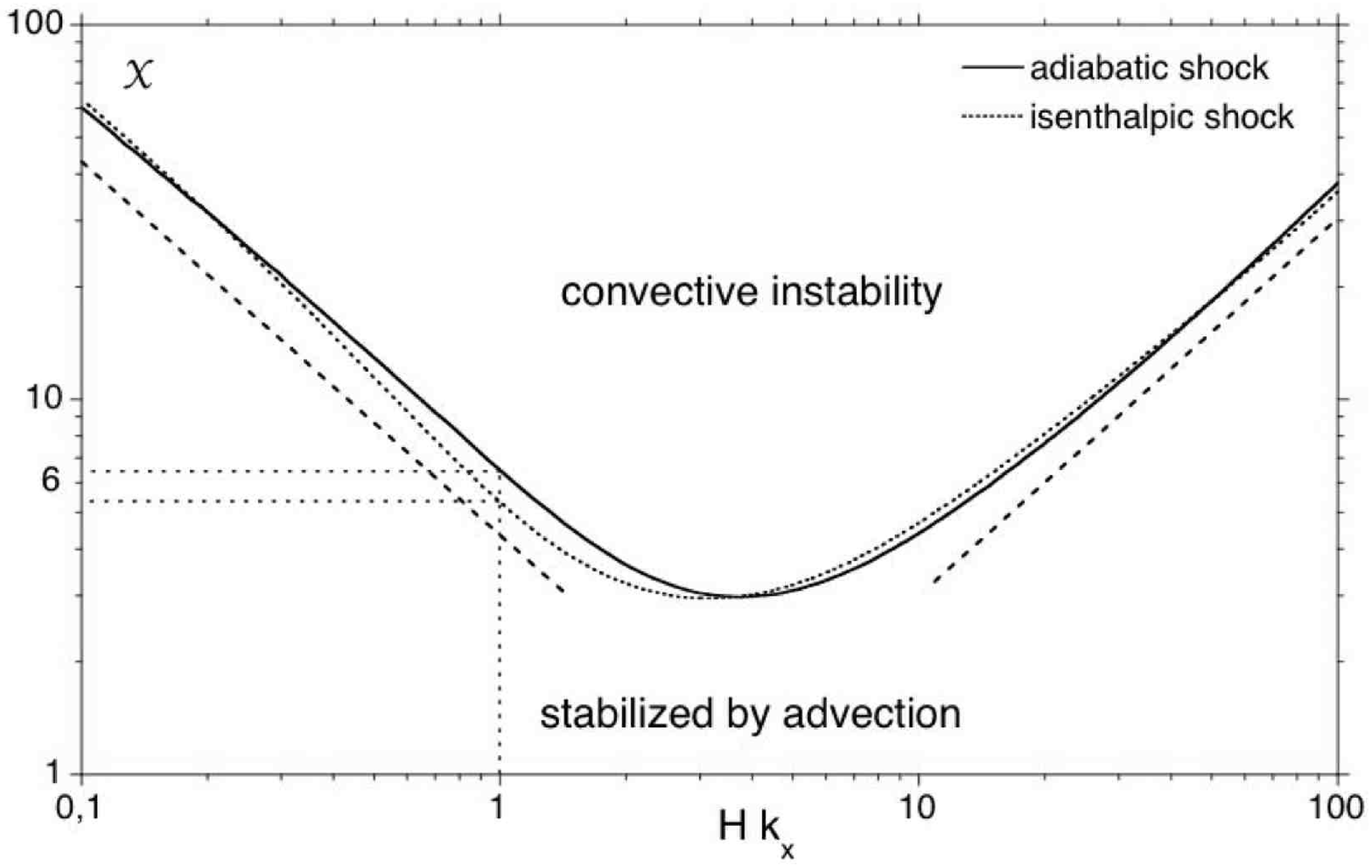}
\caption[]{Range of horizontal wavenumbers allowing the convective instability, for $\epsilon=0.1$, 
when the ratio ${\chi}$ is varied as in Fig.~\ref{figconv}. The slopes $\pm1$ are indicated as dotted 
lines for comparison. A tentative generalization of these results to a spherical
flow, in the spirit of Eq.~(\ref{extrap}), suggests that the convective instability of the $l=1$ mode in case of $H/r_{\rm sh}=0.5$
would require $\chi> 6$ (Sect.~\ref{sect_extrap}). This is indicated by
the vertical and horizontal dotted lines.}
\label{figkminmax}
\end{figure}
The numerical solution of the boundary value problem reveals that a global unstable mode grows 
exponentially with time if the advection timescale is long enough compared to the convective 
timescale. The existence of a global convective mode appears to depend directly on whether the ratio 
${\chi}$, defined by Eq.~(\ref{defT}), is above or below a certain threshold $\chi_0$.\\
As an illustration, Fig.~\ref{figconv} shows the effect of advection on the convective instability for 
$\epsilon=0.1$, where $1\le\Delta T_{\rm gain}/T_{\rm sh}\le10$ is varied so that $3<{\chi}<53$. The 
convective  growth is measured in units of $\omega_{\rm max}$, and the wavenumber in units of $H$. 
The classical convective instability is recovered in the limit ${\chi}\gg1$. The effects of advection can be 
summarized as follows:
\par(i) The growth rate is decreased compared to the maximum value of the local convective growth rate 
$\omega_{\rm max}$ (Eq.~(\ref{wbuoymax})),
\par(ii) short wavelength perturbations are stable,
\par(iii) long wavelength perturbations are also stable.\\
In Fig.~\ref{figconv}, the convective instability disappears completely for ${\chi}\le3$. The flow is then 
stable although the entropy gradient is negative.
Even when the heating coefficient $\epsilon$ is varied over three orders of magnitude, the
threshold of marginal stability always corresponds to ${\chi}_0\sim 3$, as shown in 
Fig.~\ref{figthreshTau} for an adiabatic shock (full line). This threshold is approximately insensitive to the 
loss of energy at the shock through dissociation (dashed and dotted lines).\\
The stability threshold in Fig.~\ref{figthreshTau} and in subsequent figures is obtained by solving the 
boundary value problem 
corresponding to the neutral mode (\ie $\omega=0$), as described in Appendix~D.
According to Fig.~\ref{figwavenumber}, the wavelength of the neutral mode for $\chi=\chi_0$ 
is comparable to 
$[1-3]$ times the size $H$ of the gain region, with very little influence of dissociation at the shock. A 
similar range was obtained for the first unstable mode in classical convection stabilized by viscosity 
(Sect.~\ref{Sectclassical}). \\
The range of unstable wavenumbers decreases with ${\chi}$ in a very simple way illustrated by 
Fig.~\ref{figkminmax} for $\epsilon=0.1$. When measured in units of $1/H$, the minimal and maximal 
wavenumbers are proportional to ${\chi}^{-1}$ and ${\chi}$ respectively. 

\subsection{Towards a physical understanding}

The existence of a stability threshold measured by $\chi$ can be interpreted in terms of energy. 
Approximating the entropy gradient $\nabla S\sim \Delta S/H$, the parameter $\chi^2$ 
is a measure of the potential energy in the gain region divided by the kinetic energy (\ie the inverse of the Froude number "${\rm Fr}$"):
\begin{eqnarray}
\chi&\sim&  \left({G\Delta S\over H}\right)^{1\over2}{H\over v}\sim 
\left({GH \over v^2}\right)^{1\over2}(\Delta S)^{1\over2}\propto {\rm Fr}^{-{1\over2}}.\label{roughchi}
\end{eqnarray}
The convective instability is driven by the potential energy, which is liberated by the interchange of high entropy and low entropy gas. The global instability
requires that the energy gained from the interchange is large enough to overcome the kinetic energy 
of the gas. This qualitative interpretation does not explain, however, the relatively high value 
($\chi_0\sim3$) of the threshold. Besides, the simple scaling laws measured numerically in Fig.~\ref{figkminmax} call for a simple physical mechanism, yet to be determined.

\section{Extrapolation to stalled accretion shocks in spherical geometry \label{sectappli}}

This Section aims at discussing the validity of the results of Sect.~\ref{convadv}
in a spherically symmetric flow, where convergence effects may play an important role. 
The effect of convergence is first estimated 
by a direct extrapolation of the results of Sect.~\ref{convadv}. The outcome of this 
extrapolation is then compared to the result of a numerical determination of the eigenmodes in a spherical toy model. This successful comparison suggests that the results obtained may 
keep some relevance in the more complicated context of the core-collapse problem.

\subsection{Tentative extrapolation of the parallel toy model\label{sect_extrap}}

\begin{figure}
\plotone{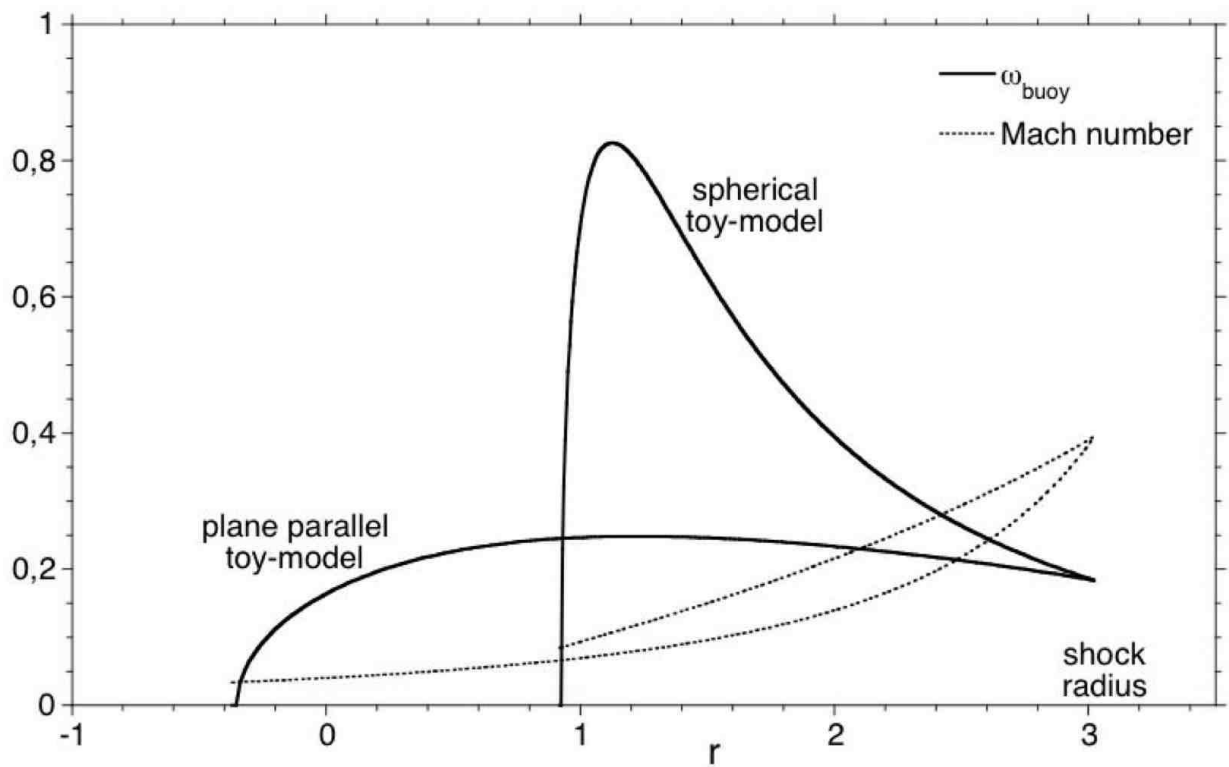}
\caption[]{
Two examples of the vertical shape of the local Brunt-V\"ais\"ala growth rate 
$\omega_{\rm buoy}(r)$ (full line) and Mach number (dotted line), in a plane parallel flow with constant 
gravity ($\chi=6.5$) and in a spherical flow ($\chi=4.1$). Distances are normalized by $c_{\rm sh}^2/G$ 
and growth rates by $G/c_{\rm sh}$, where $G$ is the strength of gravity at the shock radius.
Both flows have the same heat flux at the 
shock. Buoyancy is maximum close to the gain radius in the spherical flow. In both cases, the buoyancy 
drops abruptly to zero at the gain radius.}
\label{figwbuoyr}
\end{figure}
\begin{figure}
\plotone{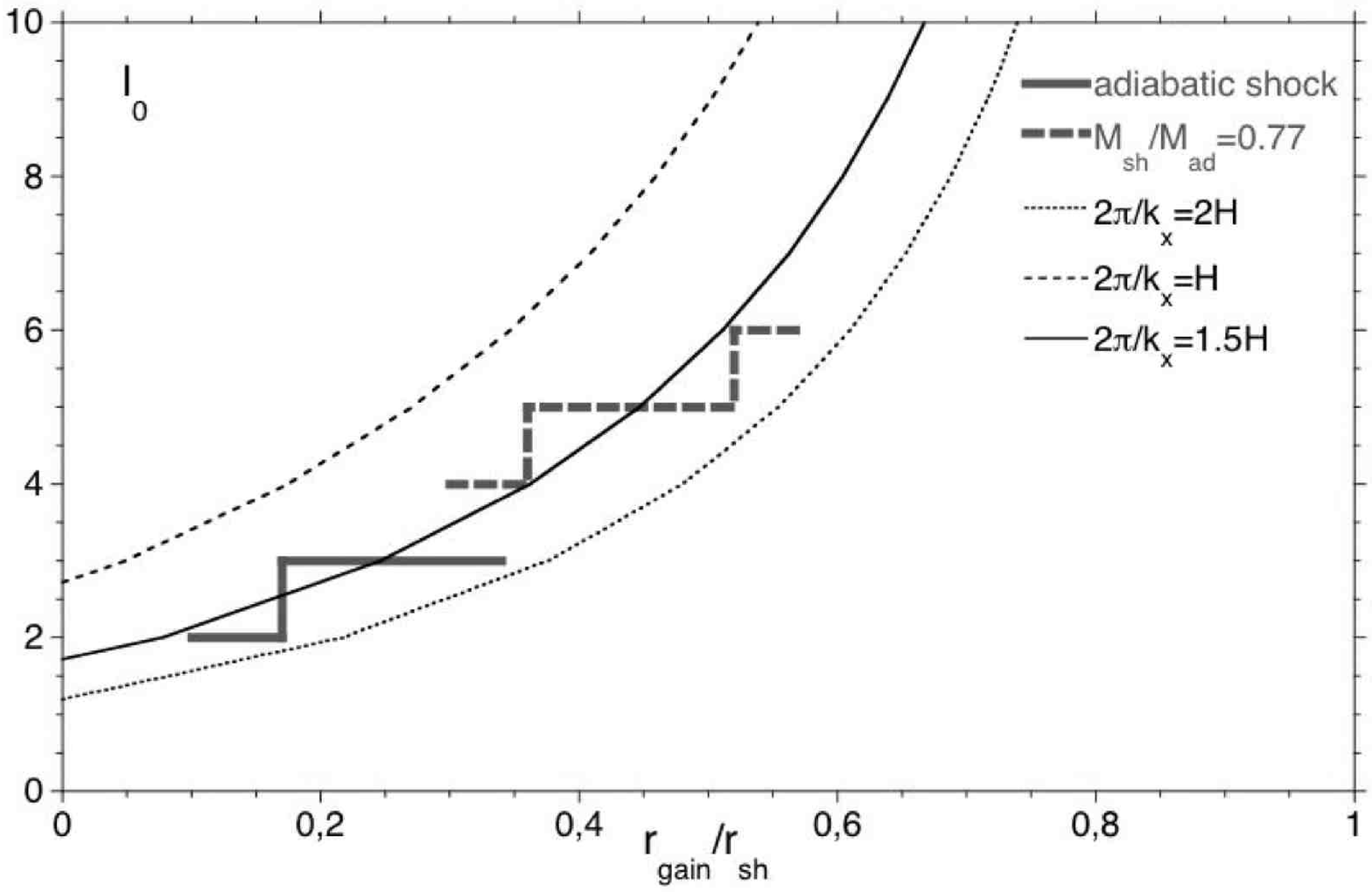}
\caption[]{
Degree $l$ of the first unstable mode in the spherical toy model for $\epsilon\le0.3$. The thin lines correspond to Eq.~(\ref{llast}) extrapolated from the parallel toy model, with $2\pi/k_x=H$ (dashed line), $2\pi/k_x=1.5H$ (full line) and $2\pi/k_x=2H$ (dotted line).The thick lines correspond to the numerical solution of the eigenmodes in a spherical gain region with an adiabatic 
shock (full line) and with dissociation (dashed line). }
\label{figlopti}
\end{figure}

The equations describing a spherical toy model and its perturbations are written in Appendix~ÊE and 
solved numerically in the next subsection. The changes introduced in spherical symmetry appear on the 
stationary flow equations: 
\begin{eqnarray}
{\cal L}&\equiv& {\LL\rho\over 1-\beta}\left\lbrack\left({r_{\rm sh}\over r}\right)^2-
\beta\left({T\over T_{\rm sh}}\right)^6\right\rbrack,\label{heatfuncsp}\\
{\p \over\p r}(\rho v r^2) &=&0,\label{eqcontsp}\\
{\p\over\p r}\left({v^2\over 2}+{c^2\over\gamma-1}-{{\cal G}M\over r}\right)&=&
{c^2\over\gamma}{\p S\over\p r}\label{eqbernsp}.
\end{eqnarray}
Neutrino heating, gravity $G\equiv {{\cal G}M/ r^2}$ and momentum $\rho v$ increase inward like 
$r^{-2}$. The radial shape of the Brunt-V\"ais\"ala frequency $\omega_{\rm buoy}(r)$ is modified 
accordingly:
\begin{eqnarray}
{\omega}_{\rm buoy}&=&
{{\cal G}M\over c r_{\rm sh}^2}
{r_{\rm sh}\over r}
\left|{v_{\rm sh}\over v}\right|^{1\over2}{(\alpha\epsilon)^{1\over2}\over (1-\beta)^{1\over2}}
\left\lbrack\left({r_{\rm sh}\over r}\right)^2-\beta\left({T\over T_{\rm sh}}\right)^6\right\rbrack^{1\over2}.
\end{eqnarray}
The critical heating leading to a postshock reacceleration is decreased by a factor $\alpha\sim 0.5$ 
due to the convergence of the flow so that $\epsilon$ is now defined as
\begin{eqnarray}
\epsilon&\equiv &-{\gamma-1\over\alpha}{{\LL}r_{\rm sh}^2\over {\cal G}M v_{\rm sh}}\ ,\\
\alpha&\equiv&1-{2r_{\rm sh}c_{\rm sh}^2\over {\cal G}M }\ .
\end{eqnarray}
According to this new definition of $\epsilon$, its estimation for a spherical flow is:
\begin{eqnarray}
\epsilon&\sim& 0.32\;\left({v_1\over7v_{\rm sh}}\right)
\left({r_{\rm sh}\over 150{\rm km}}\right)^{1\over 2} .\label{estiepsilonsph}
\end{eqnarray}
As illustrated on Fig.~\ref{figwbuoyr} for $r_{\rm gain}/r_{\rm sh}\sim0.3$, the maximum value of 
$\omega_{\rm buoy}(r)$ may be reached close to the gain radius, and may exceed its value at the 
shock by a large factor. As seen in Appendix~E, the structure of the perturbed equations (\ref{dfsp}-\ref{dksp}) is the same as in 
the parallel toy model (Eqs.~(\ref{dfp}-\ref{dkp})), the main change consisting in replacing $k_x^2$ by 
$l(l+1)/r^2$. A tentative extrapolation of the results obtained in a parallel flow may use 
a mean radius $(r_{\rm sh}+r_{\rm gain})/2$ to translate the horizontal wavenumber $k_x$ into the 
degree $l$. This gives:
\begin{equation}
l^{1\over2}(l+1)^{1\over2}\sim {k_x\over 2} (r_{\rm sh} + r_{\rm gain}) 
= k_x H\left({r_{\rm sh}\over H} - {1\over 2}\right).\label{extrap}
\end{equation}
Since the most unstable horizontal wavelength is comparable to $[1-2]$ times the vertical 
size of the gain region in the parallel flow (Fig.~\ref{figwavenumber}), the degree $l$ of the most 
unstable mode should be comparable to
\begin{eqnarray}
l^{1\over2}(l+1)^{1\over2}\sim [1-2]\pi \left({r_{\rm sh}\over H}-{1\over2}\right).\label{llast}
\end{eqnarray}
Applying in this formula $H/r_{\rm sh}\sim 0.5$ leads to $l\sim 6$ (thin full line in Fig.~\ref{figlopti}). This correspondence is certainly very crude for low degree modes. Nevertheless, it compares favorably to the results obtained numerically in spherical geometry in the next subsection (thick lines in Fig.~\ref{figlopti}). Making a similar extrapolation for the $l=1$ mode in a flow
with $H/r_{\rm sh}\sim 0.5$, in which case Eq.~(\ref{extrap}) gives $k_xH \sim 1$,
Fig.~\ref{figkminmax}  suggests that this mode should be stabilized by advection
unless ${\chi}>6$.

\subsection{Numerical solution for the eigenmodes in a spherical toy model}

\begin{figure}
\plotone{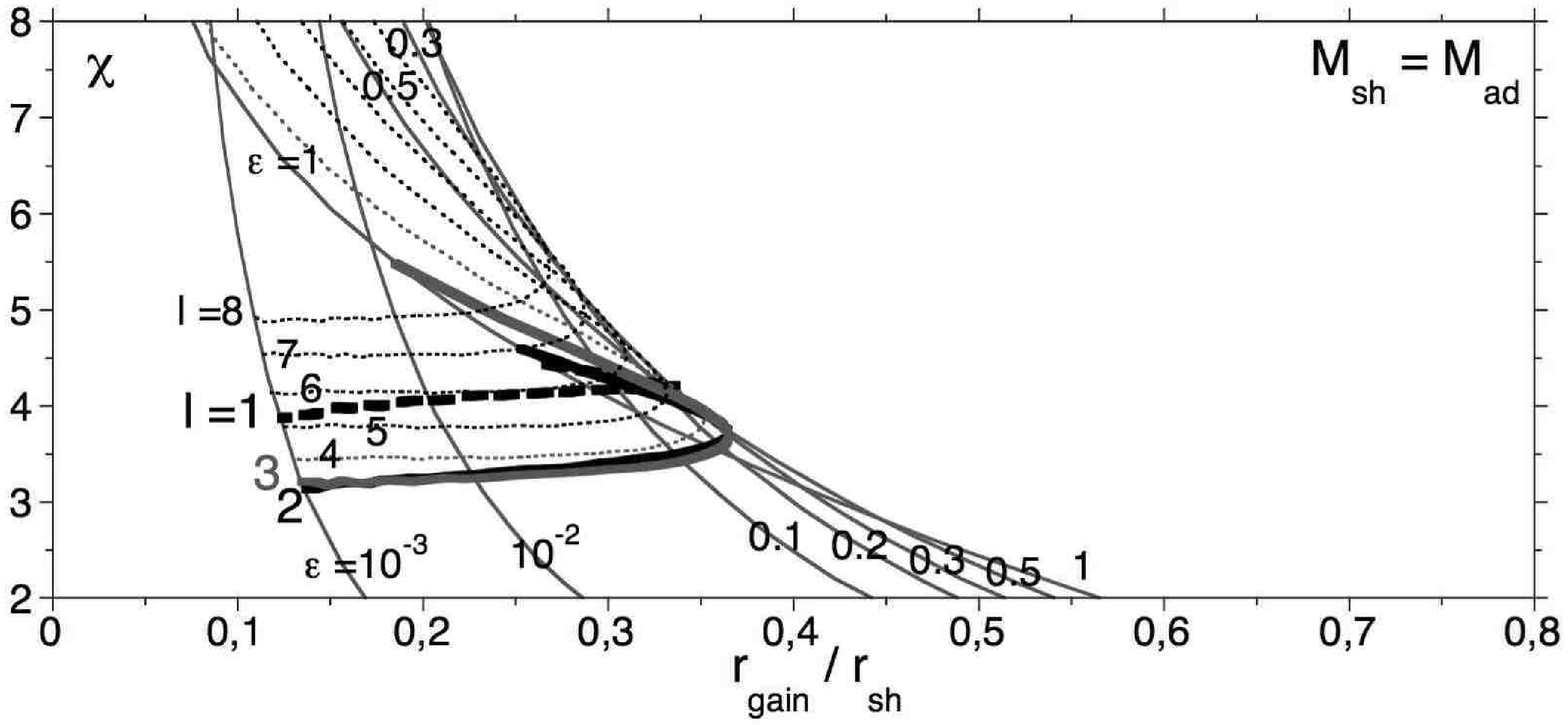}
\plotone{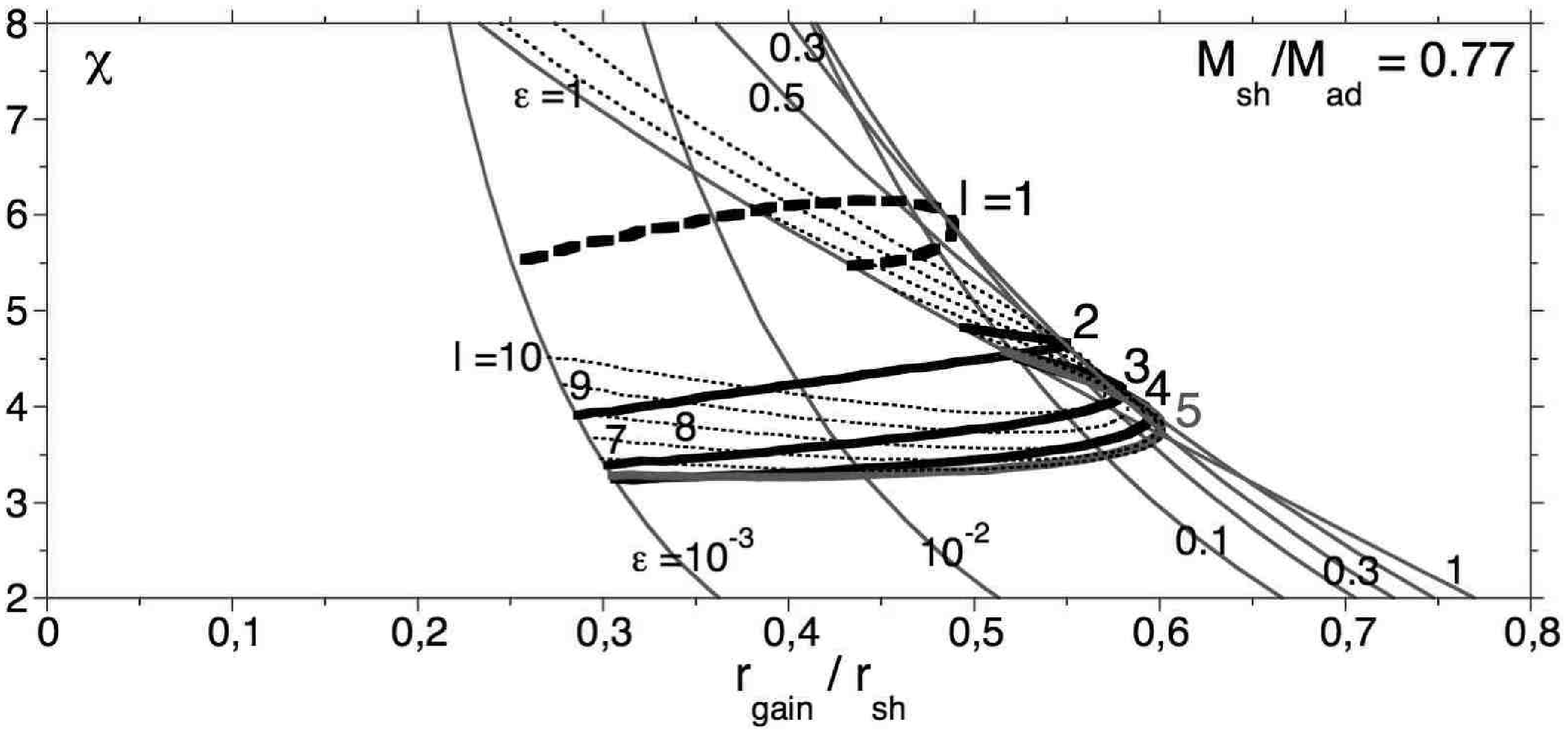}
\caption[]{Instability threshold determined from solving numerically the boundary value problem associated to a toy model in spherical geometry.
The thin full lines show the range of ($r_{\rm gain},\chi$) of the flows with a given value of the heating parameter $\epsilon$, indicated on each line, when the cooling parameter $\beta$ is varied. Nuclear dissociation is neglected in the upper plot, whereas $M_{\rm sh}/M_{\rm ad}=0.77$ in the lower plot. 
The curves labelled by $l$ (thick lines and thin dotted lines) show the stability threshold of a specific 
low degree perturbation, when the heating parameter $\epsilon$ is varied from $\epsilon=10^{-3}$ to 
$\epsilon=1$. The mode $l=1$ is associated with the thick dashed line. The most unstable modes 
correspond to $l\sim2-3$ perturbations in flows without dissociation (upper plot), and $l=4-5-6$ with 
dissociation (bottom plot). In both cases, the global stability threshold corresponds to $\chi_0\sim 3$.}
\label{figchisph}
\end{figure}

\subsubsection{Range of $r_{\rm gain}/r_{\rm sh}$ and $\chi$ in our stationary, spherical toy model}

The spherical geometry strongly limits the range of parameters ($r_{\rm gain}/r_{\rm sh},\chi$) that can 
be reached within reasonable values of neutrino heating ($\epsilon\le 1$).
Due to the geomatric dilution of neutrino heating in Eq.~(\ref{heatfuncsp}), the temperature contrast with the gain region $\Delta T_{\rm gain}/T_{\rm sh}$ is no longer an explicit parameter $\beta$ of the cooling function, because Eq.~(\ref{betat}) is now replaced by
\begin{eqnarray}
{\Delta T_{\rm gain}\over T_{\rm sh}}=
{1\over \beta^{1\over6}}\left({r_{\rm sh}\over r_{\rm gain}}\right)^{1\over3}-1.\label{betatsp}
\end{eqnarray}
It is thus more convenient to use the parameters ($\epsilon,\beta$) to define a toy model in spherical symmetry.
The thin full lines of Fig.~\ref{figchisph} show how the parameter space ($\epsilon,\beta$) can be mapped into the plane ($r_{\rm gain}/r_{\rm sh},\chi$). This 
mapping is folded near $\epsilon\sim0.3$. Note that this folding is also present in the plane parallel toy 
model, as can be deduced from Figs.~\ref{figgdr} and \ref{figtauRT}. In spherical geometry,
high values of $\chi$ can only be obtained in flows with a 
small gain radius. Even with the decelaration due to dissociation (right plot of Fig.~\ref{figchisph}), 
the flows where $r_{\rm gain}/r_{\rm sh}>0.6$ have a very moderate parameter $\chi<4$.

\subsubsection{Numerical solution of the boundary value problem}

The stability analysis of this family of flows is summarized by the thick full lines and thin dotted lines of 
Fig.~\ref{figchisph}, obtained as follows: for each heating parameter $\epsilon$ in the range 
$10^{-3}\le\epsilon\le1$, a parameter $\beta(\epsilon,l)$ is determined such that the corresponding 
flow is marginally stable with respect to perturbations of degree $l$. The value of the gain radius and 
$\chi$ of this flow is then plotted. 
\par - In both plots, the threshold for convective instability corresponds to $\chi_0\sim3$, as in plane 
parallel flows.
\par - The degree $l$ of the first unstable mode, plotted in Fig.~(\ref{figlopti}) is surprisingly close to the 
rough estimate extrapolated from the parallel toy model (Eq.~(\ref{llast})) for $\epsilon\le0.3$. 
\par - The destabilization of the mode $l=1$ is slightly easier than anticipated by the plane parallel toy 
model. The threshold $\chi_0^1$ for this mode is in the range $[4,6]$, whereas Fig.~\ref{figkminmax} 
would suggest $\chi_0^1\sim 6\pm1$. Moreover, dissociation surprisingly increases the $l=1$ 
threshold in a spherical flow, while it  would be slightly decreased according to Fig.~\ref{figkminmax}. 

\section{Comparison to supernova simulations\label{sect_simul}}

\begin{figure*}
\includegraphics[width=16.cm]{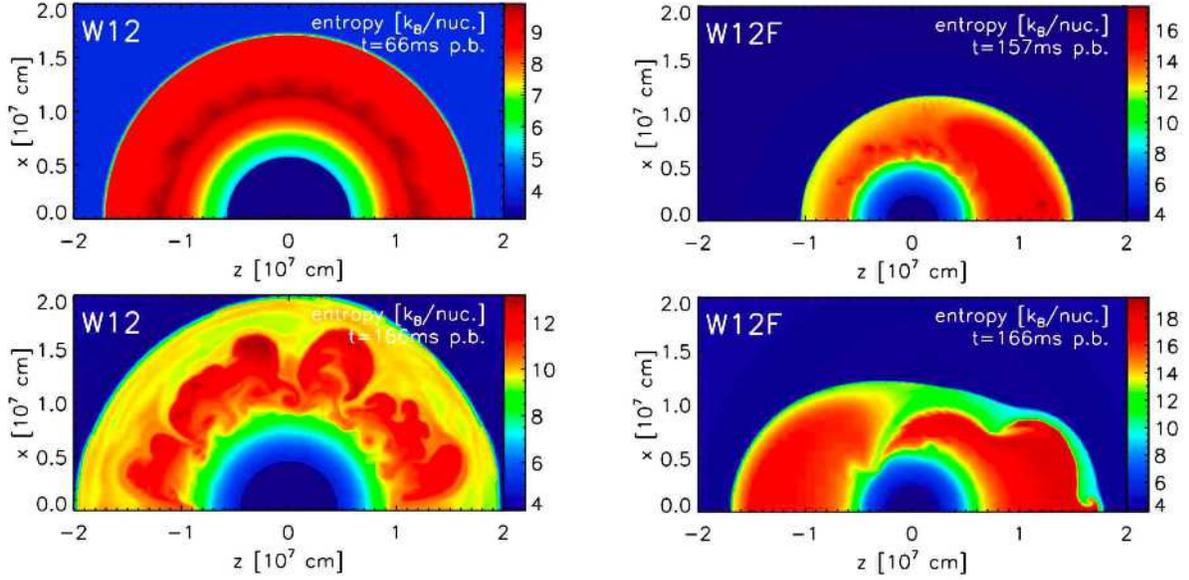}
\caption[]{Results of two hydrodynamic simulations (in 2D) of the 
post-bounce accretion phase of a stalled supernova
shock. The plots on the left show snapshots (at $t = 66\,$ms
and 166$\,$ms after bounce) from a 
model with slow neutron star contraction and $\chi = 5$.
In this case convection develops in the neutrino-heating
layer with a dominant $l = 4\,$--$\,5$ mode. The right 
plots display the situation for a more rapidly
contracting neutron star, in which case $\chi = 2.5$. 
As predicted by the analytic toy model of Sect.~\ref{sectappli}, 
convection is unable to grow for these conditions.
The visible instability of an $l=1$ mode is oscillatory 
and distinctively different from the development of 
convection in the other case. This asymmetry is interpreted 
as the consequence of a global nonradial instability of the 
accretion shock, which is of advective-acoustic nature. 
Entropy inhomogeneities created by the oblique
shock then seed the growth of Rayleigh-Taylor mushrooms
as a secondary phenomenon.}
\label{figMPA}
\end{figure*}

The perturbative analysis presented here was accompanied by
two-dimensional hydrodynamic simulations of the accretion
phase of stalled supernova shocks after core bounce. While
the details of these numerical experiments will be reported
elsewhere (Scheck et al., in preparation), we will discuss
some of the results here in order to link the conclusions
from the simplified toy model to the flow dynamics found
in more complete hydrodynamic simulations of the conditions 
in supernova cores.

The simulations were performed with the computational
setup described in Scheck et al.\ (2006). 
We used the Riemann-solver based {\em Prometheus} 
hydrodynamics code, which was supplemented by an  
approximative (grey) description of the neutrino transport.
This allowed us to follow the effects of neutrino cooling
and heating and their backreaction on the neutrino fluxes.
The simulations made use of a physical equation of state
of the stellar plasma including leptons, photons, and 
baryons. The baryonic composition is assumed to be given
by neutrons, protons, alpha particles, and
a representative heavy nucleus in nuclear statistical
equilibrium, thus ensuring the inclusion of 
nuclear photodisintegration effects. The two-dimensional 
models were computed with a polar grid and started from 
spherically symmetric post-bounce conditions of the 
collapsing core of a 15$\,M_\odot$ star, perturbing the
initial data for seeding convection. In order to compare
the numerical simulations with the analytic analysis, these
perturbations must be sufficiently small to trigger the
onset of instability in the linear regime.
The inner, high-density core of the neutron star was
replaced by a gravitating point mass and a Lagrangian 
(and thus closed) boundary at a chosen, shrinking radius
to mimic the contracting nascent neutron star. The neutrino
fluxes and mean flux energies at this boundary (which is
typically located at an optical depth much larger than 10 for 
all neutrino flavors) were prescribed as functions of time.

The behavior of a model depends on the motion of the inner
grid boundary and the size of the boundary fluxes.
When the latter are sufficiently high,
the model develops an explosion; when the luminosities are
below some critical value, no explosion can occur. The
influence of the boundary motion is more subtle. On the
one hand, it directly affects the development of nonradial
hydrodynamic instabilities in the postshock layer, a fact
which will be used below for setting up special conditions
in the supernova core. On the other hand, it determines the 
amplification of the neutrino luminosities in the settling
surface layer of the contracting neutron star. The faster
the neutron star contracts, the more it heats up by the conversion
of gravitational energy to internal energy. As a consequence,
the accretion luminosity becomes higher. This produces stronger 
neutrino heating behind the shock and thereby has a
bearing on the supernova dynamics.

The direct influence of the boundary contraction results 
from the fact that the accretion shock follows the boundary
behavior when nonradial hydrodynamic instabilities are absent 
and the neutrino luminosities 
are not close to the critical value for causing an explosion.
A more compact neutron star thus leads to a smaller shock
radius and correspondingly higher infall velocities ahead of
and behind the shock. Since the gain layer is also more
narrow, the advection timescale of the accreted gas and 
therefore the parameter $\chi$ in the gain layer is
significantly reduced. This dependence allowed us to tune
the possibility for the development of convective instability
through the chosen contraction of the inner boundary. 

Figure~\ref{figMPA} shows snapshots from two such numerical
experiments. The plots on the left result from a simulation
(Model~W12 from Scheck et al.\ 2006) with a slowly contracting 
neutron star and therefore a large value of 
$\chi_{\mathrm{W12}} = 5$, which enables the gain layer to
become convectively unstable ($r_{\mathrm{gain}}/r_{\mathrm{sh}}
\approx 0.6$ in this model at $t \ga 65\,$ms after bounce). 
As predicted by the perturbative analysis in Sect.~\ref{sectappli} (see the
case with nuclear dissociation in Fig.~\ref{figchisph}),
the most unstable and fastest growing mode is found to have 
$l = 4\,$--$\,5$ and a corresponding wavelength of about $2H$.
Note that for clean diagnostics, the power distribution of 
the initial seed must not disfavor this wavelength compared
to others (as, e.g. random zone-to-zone seed perturbations do).

The right plots display the
situation for a more rapidly contracting neutron star 
(Model W12F of Scheck et al.\ 2006), in which case we
determined $\chi \approx 2.5$. In agreement with the
analytic toy model, convection does not develop
in the first place here.
The bipolar sloshing mode and $l = 1$ deformation that 
grows instead is distinctively different from the dynamics
of the other case and not
of convective nature, but is interpreted as a nonradial accretion shock
instability that originates from 
an advective-acoustic cycle in the volume between shock
and neutron star surface. This will be further analyzed 
and discussed in a forthcoming paper (Scheck et al., in
preparation). Subsequently, convective activity is seeded
by the increasingly larger entropy perturbations that are
created in the accretion flow when it passes the deformed 
shock obliquely. This is visible in the right panels of 
Fig.~\ref{figMPA} in mushroom-like Rayleigh-Taylor fingers showing up.
Numerical experiments like these with a more ``realistic''
description of the supernova conditions therefore confirm
the conclusions drawn from the toy model (whose free 
parameters and crucial features were, of course, designed
to accommodate the physical conditions during the stalled
shock accretion phase).

We point out, however, that simulations of the full supernova
problem are complex and can involve factors which can make
it harder to unambiguously disentangle the action of different
instabilities and their properties than for the described 
specific setups. Of particular relevance in this respect
are the properties of the perturbations
that trigger the growth of convection. If the seed power of 
the most unstable convective mode is much lower than for other
wavelengths, for example, the growth of these other modes
may initially be favored. In case of large perturbation 
amplitudes on the other hand, the conditions for the linear 
regime of the analytic
analysis may not be fulfilled and buoyant bubble floating may
set in even if the linear analysis predicts convective stability
of the accretion flow. A comparison of numerical simulations with
the analytic discussion therefore requires special care.

\section{Conclusions\label{sect_conclusion}}

A toy model has been developed and studied in depth in order to understand the effect of advection 
on the linear growth of the convective instability in the gain region immediately below a stationary 
shock.
New results have been obtained through a numerical calculation of the eigenmodes and 
extrapolated to core-collapse flows. \\

\par (i) The numerical solution of the boundary value problem reveals that convection can be 
significantly stabilized by advection. The existence of a negative entropy gradient is not a 
sufficient condition for the convective instability in an advected flow.
Not only the growth rate of the fastest growing mode is diminished, 
but also the range of unstable wavelengths is modified. 
\par - The effect of advection on the convective instability is essentially governed by a single 
dimensionless parameter $\chi$, defined by Eq.~(\ref{defT}), which compares the buoyant and the 
advection timescales. 
\par - As illustrated in Fig.~\ref{figconv}, a convective mode may develop in the gain region, with a 
linear growth rate significantly slower than the local convective growth rate if the value of $\chi $ is 
moderate ($3<\chi<5$). In the plane parallel flow of our toy model, the gain region is linearly stable if 
$\chi<3$. 
\par - The minimum and maximum unstable wave numbers are directly related to $\chi$ according to 
$Hk_{\rm min}\propto 1/\chi$, $Hk_{\rm max}\propto\chi$ as shown in Fig.~\ref{figkminmax}.
\par - The horizontal wavelength of the first unstable mode when $\chi\sim 3$ is comparable to twice 
the vertical size of the gain region.\\ 

\par(ii) These results can be used as a guide to better understand the convective motions behind a 
stalled accretion shock in core collapse supernovae, and in particular their contribution to an $l=1$ 
asymmetry. The parameter $\chi$ can be measured directly in 
numerical simulations. A rough estimate, 
in Eq.~(\ref{Estichi}), suggests that $\chi$ may be close to the threshold of stabilization.
The stabilization of long wavelength perturbations, illustrated by Fig.~\ref{figconv} and 
Fig.~\ref{figkminmax}, can be a severe impediment to the development of a residual $l=1$ mode. 
The solution of the eigenvalue problem in a spherical toy model confirms the threshold 
$\chi_0\sim3$ for the convective instability, and shows examples where the mode $l=1$ is stable 
unless $\chi>4-6$. A comparison with numerical simulations for supernova conditions supports the conclusions of our toy model. \\

To what extent should we expect the threshold $\chi_0\sim3$ to be valid in the more complicated 
setup of core-collapse simulations ? Although we did not provide a quantitative explanation for this particular value, it is remarkable that the threshold seems to be affected very little ($2.6<\chi_0<3.2$) by significant changes in the flow parameters, including both the heating and cooling parameters and the photodissociation at the shock (Fig.~\ref{figthreshTau}). This threshold is also independent of the geometry; the same value applies for plane parallel and spherical conditions 
(Fig.~\ref{figchisph}). \\

We should stress again an important hypothesis of our toy model concerning the boundary condition at the gain radius: a leaking boundary condition isolates the gain region from any possible acoustic feedback from below the gain radius. The threshold $\chi_0\sim3$ thus sets the limit of a convective instability fed by the gain region alone.
Whether coupling processes taking place {\it below} the gain radius may or may not help convective motions to develop, cannot be answered by the present study. Preliminary results by Yamasaki \& Yamada (2006), showing an instability for $\chi<3$, could be interpreted as partially fed by an acoustic feedback from below the gain radius.  \\

More generally, our results do not preclude the possible destabilization of the $l=1$ mode due to coupling processes occurring below the gain radius (Blondin \etal 2003, Galletti \& Foglizzo 2005, Scheck et al. 2004, 2006, Ohnishi, Kotake \& Yamada 2006, Burrows \etal 2006). The study of such global cycles, and their possible interaction with convection, is the subject of forthcoming papers which consider a linear analysis (Foglizzo et al. 2006) as well as numerical
simulations of the nonlinear growth of nonspherical modes in the supernova core (Scheck et al., in preparation).

\acknowledgements
TF thanks Pascal Galletti for stimulating discussions. 
The authors are grateful for funding by Egide (France) 
and by DAAD (Germany) through their ``Procope" exchange program.
Support by the Sonderforschungsbereich 375 on ``Astro-Particle Physics" of the Deutsche 
Forschungsgemeinschaft is acknowledged.

\appendix

\section{Photodissociation at the shock}

The energy cost of dissociation is ${\cal E}\sim 8.8$MeV per nucleon for iron. This energy is assumed 
to be lost immediately after the shock. It affects the conservation of the Bernoulli constant across the shock as a sink term ${\cal E}$ on the post-shock side:
\begin{eqnarray}
{v_1^2\over2}+{c_1^2\over\gamma-1}={v_{\rm sh}^2\over2}+{c_{\rm sh}^2\over\gamma-1}+{\cal E}.
\label{ber}
\end{eqnarray}
The conservation of mass flux $\rho v$ and momentum flux $P+\rho v^2$ across the shock are unchanged by photodissociation. Writing $P=\rho c^2/\gamma$, these two equation are: 
\begin{eqnarray}
\rho_1 v_1&=&\rho_\sh v_\sh,\label{rov}\\
\rho_1{c_1^2\over\gamma}+\rho_1 v_1^2&=&\rho_\sh{c_\sh^2\over\gamma}+\rho_\sh v_\sh^2.
\label{rov2}
\end{eqnarray}
The classical Rankine-Hugoniot jump conditions are replaced by the following formulation, obtained after some algebra with Eqs.~(\ref{ber}-\ref{rov2}):
\begin{eqnarray}
{{\cal E}\over {v_1^2\over2}+{c_1^2\over\gamma-1}}&=&
\left(1-{\M_{\rm sh}^2\over\M_{\rm ad}^2}\right)\left(1-{\M_{\rm sh}^2\over\M_1^2}\right),
\label{paramdiss}\\
{v_{\rm sh}\over v_1}&=&{\M_{\rm sh}^2\over\M_1^2}
{1+\gamma\M_1^2\over1+\gamma\M_{\rm sh}^2},\label{v2v1}\\
{c_{\rm sh}\over c_1}&=&
{\M_{\rm sh}\over\M_1}{1+\gamma\M_1^2\over1+\gamma\M_{\rm sh}^2}
\label{T2T1}.
\end{eqnarray}
Using Eq.~(\ref{paramdiss}), we choose to parametrize the effect of dissociation by the value of the postshock Mach number $\M_{\rm sh}\le\M_{\rm ad}$. According to Eqs.~(\ref{v2v1}) and (\ref{T2T1}), $v_1/v_{\rm sh}=\gamma\M_1^2$ and $c_{\rm sh}=c_1$ if 
$\M_{\rm sh}\M_1=1/\gamma$. This rather extreme case is refered to as the ``isenthalpic shock".

\section{Expression of the perturbed heating/cooling function}

Using the fact that $\delta T/T=\delta c^2/c^2$, the heating function described by Eq.~(\ref{heatfunc}) is perturbed as follows:
\begin{eqnarray}
\delta\left({{\cal L}\over \rho v}\right)&=&-{\nabla S}{c^2\over \gamma}{\delta v_z\over v}
-{6\beta{\nabla S}_{\rm sh}\over\gamma(1-\beta)}{v_{\rm sh}\over v}c_{\rm sh}^2
\left({c\over c_{\rm sh}}\right)^{12}{\delta c^2\over c^2},\\
\delta\left({{\cal L}\over pv}\right)&=&{\gamma\over c^2}\delta\left({{\cal L}\over \rho v}\right)
-{\delta c^2\over c^2}{\nabla S},
\end{eqnarray}
In these equations, the perturbations $\delta v_z$ and $\delta c$ can be replaced by functions of $f,h,\delta S$ using Eq.~(\ref{defS}) and Eqs.~(\ref{defif}-\ref{defih}):
\begin{eqnarray}
{\delta v_z\over v}&=&{1\over 1-\M^2}\left( h-{f\over c^2}+{\delta S}\right),\\
{\delta c^2\over c^2}&=&{\gamma-1\over 1-\M^2}\left({f\over c^2}-\M^2{ h}
-\M^2{\delta S}\right).
\end{eqnarray}

\section{Boundary conditions}

\subsection{Shock boundary condition}

The boundary condition at the shock are established in this Appendix, following the conservation of mass flux, momentum flux and energy flux in the frame of the shock:
\begin{eqnarray}
\rho_{1}(v_{1}-\Delta v)&=&(\rho_{\sh}+\delta \rho_\sh)(v_{\sh}+\delta v_{\sh}-\Delta v),\\
\rho_{1}(v_{1}-\Delta v)^{2}+\rho_{1}{c_{1}^{2}\over\gamma}&=&
(\rho_{\sh}+\delta \rho_\sh)(v_{\sh}+\delta v_{\sh}-\Delta v)^{2}\nonumber\\
&&+(\rho_{\sh}+\delta \rho_\sh){(c_{\sh}+\delta c_{\sh})^{2}\over\gamma},\\
{(v_{1}-\Delta v)^{2}\over2}+{c_{1}^{2}\over\gamma-1}&=&
{(v_{\sh}+\delta v_{\sh}-\Delta v)^{2}\over2}
+{(c_{\sh}+\delta c_{\sh})^{2}\over\gamma-1}\nonumber\\
&&+{\cal E},
\end{eqnarray}
where $\delta v_{\rm sh}$ stands for the vertical component of the perturbed velocity, and quantities are measured at the position $\rsh+\Delta\zeta$. Keeping 
the first order terms, and using the defnition of $f,h$, these equations are rewritten at the 
position $\rsh$ using a Taylor expansion:
\begin{eqnarray}
\rho_{1}v_{1}h_\sh-(\rho_{\sh}-\rho_{1})\Delta v=,\nonumber\\
\Delta\zeta\left\lbrack
{\p\over\p z}\left(\rho v\right)_1
-{\p\over\p z}\left(\rho v\right)_\sh
\right\rbrack,\\
v_{\sh}^2\delta \rho_\sh+2\rho_\sh v_\sh\delta v_\sh
+{2\over\gamma}\rho_{\sh}c_{\sh}\delta c_{\sh}+\delta \rho_\sh {c_{\sh}^2\over\gamma}
=\nonumber\\
\Delta\zeta\left\lbrack
{\p\over\p z}\left(\rho v^2+\rho{c^2\over\gamma}\right)_1
-{\p\over\p z}\left(\rho v^2+\rho{c^2\over\gamma}\right)_\sh
\right\rbrack
,\\
f_\sh-(v_\sh-v_{1})\Delta v=\nonumber\\
\Delta\zeta\left\lbrack
{\p\over\p z}\left({v^2\over2}+{c^2\over\gamma-1}\right)_1
-{\p\over\p z}\left({v^2\over2}+{c^2\over\gamma-1}\right)_\sh
\right\rbrack,
\end{eqnarray}
The local gradients in these three equations are computed from Eqs.~(\ref{eqcont}), (\ref{eqbern}), (\ref{eqS}) describing the stationary flow,
\begin{eqnarray}
{\p\over\p z}\left(\rho v\right)&=&0,\\
{\p\over\p z}\left(\rho v^2+\rho{c^2\over\gamma}\right)&=&-G\rho,\\
{\p\over\p z}\left({v^2\over2}+{c^2\over\gamma-1}\right)&=&{{\cal L}\over\rho v}-G.
\end{eqnarray}
We obtain:
\begin{eqnarray}
h_\sh&=&\left({1\over v_{\sh}}-{1\over v_{1}}\right)\Delta v,\label{boundg}\\
{\delta S_\sh\over\gamma}&=&
{\Delta\zeta\over c_\sh^2} \left\lbrack {{\cal L}_1\over\rho_1 v_1}
-{{\cal L}_\sh\over\rho_\sh v_\sh}-G\left(1-{v_\sh\over v_1}\right)\right\rbrack\nonumber\\
&&-{v_1\Delta v\over c_\sh^2}\left(1-{v_\sh\over v_1}\right)^2,\\
f_\sh&=&(v_{\sh}-v_{1})\Delta v+\Delta\zeta\left({{\cal L}_1\over\rho_1 v_1}
-{{\cal L}_\sh\over\rho_\sh v_\sh}
\right).
\end{eqnarray}
The assumption that ${\cal L}_1\ll {\cal L}_\sh$ leads to Eqs.~(\ref{fsh}), (\ref{hsh}) and (\ref{Ssh}).\\
$\delta K$ is rewritten using its definition (Eq.~\ref{defK1}) and the transverse
component of the linearized Euler equation:
\begin{eqnarray}
\delta K=k_x^2f-\omega k_x\delta v_x.\label{Kgen}
\end{eqnarray}
The transverse velocity immediately after the shock is deduced from the conservation of the tangential 
component of the velocity, in the spirit of Landau \& Lifschitz (1989).
\begin{eqnarray}
(\delta v_x)_\sh&=&(v_{1}-v_{\sh})ik_x\Delta \zeta,
\label{dvt}.
\end{eqnarray}
Finally, Eq.~(\ref{Ksh}) is deduced from Eq.~(\ref{Kgen}), with Eqs.~(\ref{dvt}) and (\ref{fsh}).\\
The vorticity $\delta w_y$ produced by the perturbed shock, deduced from Eqs.~(\ref{defK1}), (\ref{Ssh}) 
and (\ref{Ksh}), is independent of heating:
\begin{eqnarray}
(\delta w_y)_{\rm sh}=-{ik_x\over v_{\rm sh}}\left(1-{v_{\rm sh}\over v_1}\right)
\left\lbrack (v_1-v_{\rm sh})\Delta v+G\Delta\zeta\right\rbrack.
\end{eqnarray}

\subsection{Lower boundary condition}

Establishing the leaking boundary condition requires to identify the acoustic content of the perturbation $f$ as it reaches the lower boundary. For this purpose, we use the classical decomposition into acoustic and advected perturbations (\ie Landau \& Lifschitz 1989) in an adiabatic, uniform flow. The perturbation $f_S$ associated with an advected entropy perturbation 
$\delta S$ such that $\delta K=0$, and the perturbation $f_K$ associated to $\delta K$ with 
$\delta S=0$ are deduced from the differential system (\ref{dfp}-\ref{dkp}), in which ${\cal L}\equiv 0$ (adiabatic flow) and $\dd/\dd z\equiv i\omega/v$ (advected perturbations):
\begin{eqnarray}
f_S&=&{1-\M^2\over1-\mu^2\M^2}c^2{\delta S\over\gamma}\ ,\label{fS}\\
f_K&=&{\M^2(1-\mu^2)\over1-\mu^2\M^2}{\delta K\over k_x^2}\ .\label{fK}
\end{eqnarray}
Both are associated with the same wave number $k_z^0$ of advected perturbations:
\begin{eqnarray}
k_z^0={\omega\over v}.\label{kKS}
\end{eqnarray}
Pressure perturbations $f_\pm$ correspond to the solution of the differential system (\ref{dfp}) to 
(\ref{dkp}) with ${\cal L}=0$, $\delta S=0$ and $\delta K=0$. The longitudinal wavenumber $k^\pm$ of 
acoustic perturbations is equal to
\begin{equation}
k^\pm\equiv{\omega\over c}{\M \mp \mu\over1-\M^{2}},\label{kpm}
\end{equation}
where the sign of $\mu$ is defined such that Im$(k^+)<0$ when Real$(\mu^2)<0$ (evanescent wave), 
and Real$(k^+)<0$ when Real$(\mu^2)>0$ (downward propagation). The components $f^\pm$ are 
deduced from the values of $f,h$ and Eqs.~(\ref{fS})--(\ref{fK}):
\begin{eqnarray}
f^{\pm}&=& {1\over2}f \pm {{\cal M}c^2\over2\mu}(h+\delta S)-
{1\pm\mu\M\over2}\left(f^S
\pm{f_K\over\mu\M}\right)\label{fpmgen}\ ,\\
h^{\pm}&=&\pm {\mu\over\M}{f^{\pm}\over c^2}\ .\label{gfac}
\end{eqnarray}
The leaking boundary condition at the gain radius corresponds to the absence of an acoustic flux from below the gain radius: $f^-(z_{\rm gain})=0$.

\section{Boundary value problem satisfied by the neutral mode $\omega=0$}

The differential system (\ref{dfp}-\ref{dkp}) is singular for $\omega=0$.
The linearized equations for $\omega=0$ are simplest when using $f$, $h$, $\delta S$, $\delta v_x$ 
instead of $f$, $h$, $\delta S$, $\delta K$. The Euler equation in the transverse direction 
and the definition of $\delta K$ lead to:
\begin{eqnarray}
v\delta w_y&=&ik_x{c^2\over\gamma}\delta S-ik_x f\ ,\\
\delta K&=&k_x^2f\ .
\end{eqnarray}
The differential system is thus:
\begin{eqnarray}
{\p f\over\p \z}&=&\delta\left({{\cal L}\over \rho v}\right),\\
{\p  h\over\p \z}&=&-{ik_x \delta v_x\over v},\\
{\p\delta S\over\p \z}&=&\delta\left({{\cal L}\over pv}\right),\\
{\p  \delta v_x\over\p \z}&=&{-ik_x \over v(1-\M^2)}\left\lbrace
f-v^2{  h}-
{c^2\over\gamma}\left\lbrack1+(\gamma-1)\M^2\right\rbrack\delta S\right\rbrace.
\label{dvz}
\end{eqnarray}
On the shock surface, the boundary conditions are measured for a shock displacement $\Delta\zeta$:
\begin{eqnarray}
f_{\rm sh}&=&-{\Delta \zeta}
\frac{c_{\rm sh}^{2}}{\gamma} \nabla{S} , \\
 h_{\rm sh}&=&0 ,\\
\delta S_{\rm sh}&=&-\Delta \zeta\left\lbrack 
 \nabla S+
\left(1-{v_{\rm sh}\over v_{1}}\right){\gamma\over c_{\rm sh}^{2}}\nabla \phi
\right\rbrack,\\
(ik_x\delta v_x)_{\rm sh}&=&
-k_x^2 (v_1-v_{\rm sh})\Delta\zeta\ .
\end{eqnarray}
The lower boundary condition in the adiabatic part of the flow (${\nabla S}=0$) is determined by 
remarking that
\begin{eqnarray}
{\p^2\delta v_x\over \p \z^2}-{k_x ^2\delta v_x\over 1-\M^2}=0 \ .
\end{eqnarray}
The evanescent solution when $z\to-\infty$ is selected by imposing:
\begin{eqnarray}
{\p\delta v_x\over\p \z}-{k_x \delta v_x\over (1-\M^2)^{1\over2}}=0\ .\label{clvz}
\end{eqnarray}
The continuity of $f,h,\delta S,k_xv_x$ at the lower boundary implies the continuity of 
$\p \delta v_x/\p z$ according to Eq.~(\ref{dvz}). The boundary condition (\ref{clvz}) is thus continuous 
accross the gain radius (only $f$ and $\delta S$ have discontinuous derivatives across the gain 
radius).

\section{Boundary value problem in a spherical geometry}

The toy model in spherical geometry resembles the one in Cartesian geometry, 
the main difference being the radial dependence of gravity and neutrino heating. 
Rather than rewritting all the flow equations replacing $z$ by $r$, we only note 
in this Appendix those which are modified by geometrical factors. 
The gain region is located between a stationary shock at a radius $r_{\rm sh}$ 
and a gain radius $r_{\rm gain}$.

\subsection{Differential system ruling the evolution of perturbations}

The definition of $\delta K$ is the same as in Eq.~(5) of F01:
\begin{eqnarray}
\delta  K&\equiv& r^2v.\nabla\times\delta w+{l(l+1)c^2\over\gamma}
\delta S.\label{defKsph}
\end{eqnarray}
If $\omega\ne 0$, the perturbed equations are as follows:
\begin{eqnarray}
{\p f\over\p r}&=&{i\omega v\over 1-\Mc}\left\lbrace
 h -{f\over c^2} 
 +
\left\lbrack\gamma-1+{1\over\Mc}\right\rbrack{\delta S\over\gamma}
 \right\rbrace+\delta\left({{\cal L}\over \rho v}\right),\label{dfsp}
\\
{\p{  h}\over\p r}&=&{i\omega\over v(1-\Mc)}\left\lbrace
\frac{\mu^{2} }{c^{2}} f
 -\Mc {  h}
- \delta S\right\rbrace+{i\delta K\over\omega r^2v},\label{dhsp}
\\
{\p \delta S\over\p r}&=&{i\omega\over v}\delta S
+\delta\left({{\cal L}\over pv}\right),\label{dssp}
\\
{\p\delta K\over \p r}&=&{i\omega\over v}\delta K
+l(l+1)\delta\left({{\cal L}\over \rho v}\right),\label{dksp}\\
\mu^2&\equiv&1-{l(l+1)c^2\over\omega^2r^2}(1-\Mc).
\end{eqnarray}

\subsection{Boundary conditions in the spherical toy model}

The boundary conditions for $\delta S_{\rm sh}$ and $\delta K_{\rm sh}$ include the following spherical 
corrections:
\begin{eqnarray}
{\delta S_{\rm sh}\over\gamma}&=&-\Delta \zeta
\left\lbrack {\nabla S\over\gamma}+
\left(1-{v_{\rm sh}\over v_{1}}\right){{\cal G}M\over r_{\rm sh}^2c_{\rm sh}^{2}}\left(1
-4{v_{\rm sh}\over v_1}{r_{\rm sh}v_1^2\over 2{\cal G}M}\right)
\right\rbrack\nonumber\\
&&-{ v_1\Delta v\over c^{2}}\left(1-{v_{\rm sh}\over v_1}\right)^{2}, \label{Sshsp}\\
\delta K_{\rm sh}&=&
-l(l+1)\Delta \zeta
\frac{c_{\rm sh}^{2}}{\gamma} \nabla{S}_{\rm sh} ,\label{Kshsp}
\end{eqnarray}
The leaking boundary condition is a direct extrapolation of the boundary condition obtained in Cartesian geometry:
\begin{eqnarray}
{\mu\over \M}{f\over c^2} -  h-\left(\gamma+{\mu\over\M}{1-\M^2\over1+\mu\M}\right)
{\delta S\over\gamma}
\nonumber\\
+{1-\mu^2\over1+\mu\M}{\delta K\over l(l+1)c^2}=0\ .\label{leakingsph}
\end{eqnarray}

\subsection{Neutral mode $\omega=0$ in a spherical flow}

In a spherical flow, $ik_xv_x$ is replaced by the quantity $\delta A$ associated to the divergence of the 
transverse velocity perturbation:
\begin{eqnarray}
\delta A\equiv {r\over\sin\theta}\left\lbrack
{\p\over\p\theta}(\sin\theta\delta v_\theta)+{\p\delta v_\varphi\over\p\varphi}\right\rbrack.
\end{eqnarray}
The Euler equation in the transverse directions and the definition of $\delta K$ lead to:
\begin{eqnarray}
v\delta w_\varphi&=&{c^2\over\gamma r}{\p\delta S\over\p\theta}-{1\over r}{\p f\over\p\theta}\ ,\\
v\delta w_\theta&=&-{c^2\over\gamma r\sin\theta}{\p\delta S\over\p\varphi}
+{1\over r\sin\theta}{\p f\over\p\varphi}\ ,\\
\delta K&=&l(l+1)f\ .
\end{eqnarray}
The differential system is as follows:
\begin{eqnarray}
{\p f\over\p r}&=&\delta\left({{\cal L}\over \rho v}\right),\\
{\p  h\over\p r}&=&-{\delta A\over r^2 v},\\
{\p\delta S\over\p r}&=&\delta\left({{\cal L}\over pv}\right),\\
{\p\delta A\over\p r}&=&{l(l+1) \over v(1-\M^2)}\left\lbrace
f-v^2{  h}-{c^2\over\gamma}\left\lbrack1+(\gamma-1)\M^2\right\rbrack\delta S\right\rbrace.
\end{eqnarray}
The spherical corrections for the boundary conditions are:
\begin{eqnarray}
f_{\rm sh}&=&-{\Delta \zeta}
\frac{c_{\rm sh}^{2}}{\gamma} \nabla{S}\ , \\
 h_{\rm sh}&=&0\ ,\\
\delta S_{\rm sh}&=&-\Delta \zeta
\left\lbrack \nabla S+
\left(1-{v_{\rm sh}\over v_{1}}\right){\gamma {\cal G}M\over r_{\rm sh}^2c_{\rm sh}^{2}}
\left(1-4{v_{\rm sh}\over v_1}{r_{\rm sh}v_1^2\over 2{\cal G}M}\right)
\right\rbrack\ ,\\
\delta A_{\rm sh}&=&
-l(l+1) (v_1-v_{\rm sh})\Delta\zeta.
\end{eqnarray}
The lower boundary condition is determined by choosing the evanescent solution when the flow 
gradients are neglected
\begin{eqnarray}
{\p\delta A\over \p r}
-{l^{1\over2}(l+1)^{1\over2}\over r}{\delta A\over 1-\M^2}=0\ .
\end{eqnarray}

\end{document}